\documentstyle[preprint,aps,amsmath,amsfonts,latexsym,amssymb,bbold,graphicx,rotating]{revtex4b1}
\begin{document}
\title{Absorption Cross Section of Scalar Field in Supergravity
Background}
\author{R. Manvelyan$^{1}$\footnote{e-mail:
manvel@phys.uni-kl.de, on leave from Yerevan Physics Institute},
H. J. W. M\"{u}ller-Kirsten$^{1}$\footnote{e-mail:
mueller1@physik.uni-kl.de}, J.--Q.
Liang$^{1,2}$\footnote{e-mail:jqliang@physik.uni-kl.de,\;
jqliang@mail.sxu.edu.cn}
and Yunbo Zhang$^{1,2,3}$\footnote{e-mail:ybzhang@physik.uni-kl.de,\;
ybzhang5@yahoo.com}}
\address{1. Department of Physics, University of Kaiserslautern, D-67653 Kaiserslautern, Germany\\
2. Department of Physics and Institute of Theoretical Physics,
Shanxi University, Taiyuan, Shanxi 030006, China\\
3. Laboratory of Computational Physics, Beijing Institute of
Applied Mathematics and Computational Mathematics,
Bejing 100 088, China}

\maketitle

\begin{abstract}
It has recently been shown that the equation of motion
of a massless scalar field in the background of some
specific $p$ branes can be reduced to a modified Mathieu equation.
In the following the absorption rate
of the scalar by a $D3$ brane in ten dimensions is calculated in terms
of modified Mathieu functions of the first kind, using
standard Mathieu coefficients.  The relation of the
latter to Dougall coefficients
(used by others) is investigated. The $S$--matrix obtained in
terms of modified Mathieu functions of the first kind
is easily evaluated if known rapidly convergent low energy
expansions of these in terms of products of Bessel
functions are used. Leading order terms, including the interesting
logarithmic contributions, can be obtained analytically.

\end{abstract}

\vspace{0.2cm}

\vspace{0.2cm}

\section{Introduction}
\label{sec:I}

Recently the equations of motion
of several cases of massless scalar fields propagating
in a supergravity background describing $p$--brane solitons
have been shown to be reducible to a
Schr\"odinger--like equation with a singular potential
and hence to a modified Mathieu
equation, so that various aspects, such as absorption probabilities,
become exactly calculable\cite{1,2},
which by AdS/CFT correspondence may
yield information on correlation
functions in a related
world volume effective field theory.
The singular potential appearing in the
coefficients of the metric is
in the case of the $D3$--brane the
Coulomb potential in $6$ spatial dimensions. In view of the fact, that
very few such exactly solvable cases are known
and that a Mathieu--type equation arises in a
number of such problems as a result of the
invariance of the wave equation under
various diagonal dimensional reductions on the
world volume,
these theories are of exceptional importance and deserve
to be studied in full detail. The two recent
investigations \cite{1,2} of the absorption of partial
waves of a massless scalar field by $D3$ branes
in 10 dimensions \cite{1} and by a dyonic string in
six dimensions (or a $D1/D5$ brane
intersection in 10 dimensions or extremal
2--charge black hole in 5 dimensions or $M2/M5$ brane
intersection in 11 dimensions)\cite{2}
study the resulting
modified Mathieu equation in terms of expansion coefficients
introduced by Dougall\cite{3} in 1916 and present very few
details. It is not possible to follow the calculations of these
papers without extensive work of one's own,
which is made even more difficult by singularities
of expansion coefficients that require
additional attention.
These studies are, in fact, complicated applications of modified Mathieu
functions, which, in our opinion become even
more complicated if instead of standard Mathieu
coefficients, i.e. those in modern texts,
the coefficients of Dougall are used, for the
calculation of which the authors of ref.\cite{1}
developed in addition their own algorithm.
The significance of the Mathieu equation in such
contexts can also be seen from a different
angle since the equation occurs also
as the appropriately transformed
small fluctuation equation in the study of Born--Infeld theory
in the bosonic light--brane approximation with
only an electric field ${\bf E} = -{\bf \nabla}\phi$ in
$p=3$ dimensions and the remaining components of the
vector potential as massless scalar fields, reduced to only one
field $y$ in the simplest case.
One can show that finite energy
configurations of these fields independent
of one another are not stable, but
their combination with appropriate boundary condition\cite{4,4.1}
(equivalent to the Dirichlet boundary condition) is (with supersymmetry)
a BPS state with corresponding Bogomol'nyi equation.
Investigating the stability of this
configuration, i.e. the $D3$--brane, with respect to
transverse fluctuations of both the throat or fundamental string
and the brane,
one again arrives at
an equation with the singular potential $1/r^4$ \cite{4,4.2}
which can be converted into a modified Mathieu equation\cite{5,6}.
Thus in each of these cases
a Schr\"odinger--like equation is obtained with the
singular potential $1/r^4$.  Such potentials have been the
subject of investigation 30 years ago\cite{6.1} and were motivated by the
lack of understanding of weak interactions at that time. Thus
the potential $1/r^4$ and the associated scattering problem
had also been investigated, and various forms
of the $S$--matrix had been given
\cite{5,6,7,8} in terms of modified Mathieu
functions or related functions for which -- at the latest
since the publication of refs.\cite{10} and \cite{11} --
widely used definitions and notations exist.

In view of the scarcity of fully solvable examples of
theories on a supergravity background we consider it worthwhile 
to reexamine the case of the propagation of
a massless scalar field in the presence of a 3--brane by using
modified Mathieu functions with
standard Mathieu coefficients and the $S$--matrix evaluated in terms of these.
In our opinion these calculations are more
transparent than those using Dougall coefficients and are easier
to follow with reference to modern
literature on the subject.
In view of the complexity of
the calculations, due also to the fact that
later iterations contribute to earlier lower order
terms, we present these in some detail.  Our
presentation below should therefore also enable others to
follow the reasoning, and this particularly since leading
order terms can be understood without resorting to
numerical methods.

In the following we first formulate the semiclassical gravity problem
and reduce it to the modified Mathieu equation.
We do not rederive the $S$--matrix, but recapitulate
in Appendix A the main steps in the derivation, and in
particular some steps that have not been written out
explicitly in ref.\cite{6}, this being the prime reference
on which our considerations are based. We then
consider  briefly the gauge field theory approach in a simplified
Born--Infeld version
to demonstrate how this also leads to the Mathieu
equation. Following this
we consider
the Floquet exponent
associated with Mathieu functions and show how this
has to be calculated in singular cases (such as the
the cases to be considered here and in the
S--wave case
already in the dominant approximation).
 The calculation of coefficients of
series expansions of modified Mathieu functions
is then considered and the Dougall coefficients used in
refs.\cite{1,2} are compared with ordinary, i.e. standard, Mathieu
coefficients as in ref.\cite{10}.  We
calculate examples in singular and asymptotic
cases (the latter being those that permit
one to ignore the singularities of early coefficients).
Higher order contributions are obtained with
Mathematica.  We show that Dougall coefficients are
more difficult to obtain than ordinary
coefficients -- an observation that may explain
why Dougall did not evaluate any of his own
coefficients in his work of 1916.
We then evaluate the relevant quantities appearing in the
$S$--matrix and hence the absorption probabilities and
cross sections. Where comparable, our results can be seen
to agree with those of ref.\cite{1}.
The treatment presented below makes full use of the
well established theory of the Mathieu equation
and can therefore point the way to explore
other aspects, such as application to
double--centered $D3$ branes and to higher
energies which have been discussed recently \cite{11.0}.

\section{The scalar field in the $D3$--brane metric}
\label{sec:II}

The supergravity background for an extremal $Dp$--brane in the
10--dimensional type IIB theory
is \cite{11.1,11.2,11.3}
\begin{equation}
ds^2=\frac{1}{\sqrt H}(-dt^2+dx^2_{\parallel})+\sqrt H dx^2_{\perp}
\label{1}
\end{equation}
where ($r$ being the radial coordinate in the $SO(5)$ symmetric
space orthogonal to the branes)
\begin{equation}
dx^2_{\parallel}=\sum^p_{i=1}dx^2_i,\;\;\; 
  dx^2_{\perp}=dr^2+r^2d\Omega^2_{(8-p)}
\label{2}
\end{equation}
and the harmonic function $H$ is given by
\begin{equation}
H=1+\frac{R^{(7-p)}}{r^{(7-p)}}, 
\label{3}
\end{equation}
For $p=3$, the case
of interest here,
i.e. the $D3$ brane coupled to the 4--form
RR--potential\cite{11.3,11.31},  $R$ with $R^4=4\pi g_sN\alpha^{{\prime}^2}$
($g_s$ the string coupling and $N$ the number
of $D3$ branes) is the
 radius of $S^5$ and $AdS_5$ in the socalled 
decoupling limit in which one
obtains a duality between ${\cal N}=4$
$U(N)$ supersymmetric Yang--Mills theory in 4 dimensions
and string theory in the near horizon $AdS_5\times S^5$
background \cite{11.4,11.5}. As pointed out in ref.\cite{1},
for a comparison of considerations in terms of supergravity and those in
terms of $D$--branes, one  is interested in the domain of  
small $\omega R$, where $\omega$ is the energy of the field incident
on the brane.

For a massless scalar fluctuation field $\phi$
around the dilaton field $\Phi$ given by\cite{11.1}
$$
e^{\Phi}= H^{(3-p)/4}(r)
$$
(which is constant for $p=3$) in the background
of this metric, the equation
of motion is
\begin{equation}
\frac{1}{\sqrt{g}}\partial_{\mu}\sqrt{g}g^{\mu\nu}\partial_{\nu}\phi=0
\label{4}
\end{equation}
After separation of the $S^5$ harmonics $Y(\theta_i)$,
in particular the Gegenbauer polynomial $C_l(\cos\theta)$,
where $x=r\cos\theta$, and 
a factor $e^{i\omega t}$ the radial
wave function $\psi_l(r)=y(r)/r^{\frac{5}{2}}$
of the $l$--th partial wave
of energy $\omega$ of the scalar field $\phi$ is found to satisfy
\begin{eqnarray}
&&\bigg[\frac{1}{r^5}\frac{\partial}{\partial r}
\bigg(r^5\frac{\partial}{\partial r}\bigg)
-\frac{l(l+4)}{r^2}
+ \omega^2+\frac{\omega^2R^4}{r^4}\bigg]\psi_l(r)=0,\nonumber\\
&&\bigg[\frac{\partial^2}{\partial r^2}+ \omega^2+\frac{\omega^2R^4}{r^4}
-\frac{(l+\frac{3}{2})(l+\frac{5}{2})}{r^2}\bigg]y=0
\label{5}
\end{eqnarray}
We see that for $R^4\neq 0$ this is the equation of an attractive
singular potential with coupling constant $g^2_0=\omega^2R^4$.
 For $\omega^2>0$
an incident wave allows both transmitted and reflected waves, and from
the ratio of coefficients one can determine the $S$--matrix. 
It is convenient to make the substitutions
\begin{equation}
y=r^{1/2}\phi(r),\;\;\;r=\gamma e^z,\;\;\; {\gamma}=g_0/h,
\;\;\; h^2=\omega g_o=\omega^2R^2,\;\;\; \lambda = (l+2)^2,
\label{6}
\end{equation}
which convert the range of $r$ from $0$ to $\infty$
to that of $z$ from $-\infty$ to $+\infty$. 
The equation thereby becomes the modified Mathieu equation
\begin{equation}
\frac{d^2 \phi}{dz^2}+\bigg[2h^2\cosh2z - \lambda\bigg]\phi=0
\label{7}
\end{equation}

In view of the principal interest in the relation
of our semiclassical gravity consideration with the
superconformal limit of the dual theory in the near--horizon
domain, we are here interested in waves of low
energy, i.e. of small $\omega$, and
so in solutions of the modified Mathieu equation
around $h^2=0$. 
The modified Mathieu equation allows series expansions
of this type in terms
of exponential, hyperbolic and cylindrical functions and 
(surprisingly) in each of the cases with
the same coefficients $c^{\nu}_{2r}(h^2)$
where $\nu$ is the Floquet exponent and the subscript $r$
a positive or negative integer
or zero (not to be confused with the radial coordinate).
The solutions in terms of exponentials are written
$Me_{\nu}(z,h^2)$, those in terms of hyperbolic functions
$\cosh$ and $\sinh$ $Mc_{\nu}$ and $Ms_{\nu}$. The solutions
of the $i$--th kind are those in terms
of cylindrical functions and are written $M^{(i)}_{\nu}(z,h^2)$ where
$ i=1,2,3,4 $ correspond respectively to expansions in terms
of Bessel, Neumann or Hankel$^{(1,2)}$ functions.
The series of $M^{(i)}_{\nu}(z,h^2)$
converge uniformly only for $|\cosh z|>1$, whereas the series of
$Me_{\nu}(z,h^2)$ is uniformly convergent for all finite complex values
of $z$ as shown in ref.\cite{10}. 
Since $r=0$ corresponds to $z=-\infty$, the $S$--matrix is obtained
by continuing the solution at $z=-\infty$ to
$z=\infty$. This means that a solution $M^{(3)}_{\nu}$ has to be
continued, via matching to $Me_{\pm \nu}$ (across the domain $|z|<1$),
to a linear combination of  
 $M^{(3)}_{\nu}$ and $M^{(4)}_{\nu}$ at $+\infty$. 
A  few main steps of this calculation are given in Appendix A. 
The expression for the $S$--matrix finally obtained
is
\begin{equation}
S=\frac{R^2-1}{R^2-e^{-2i\pi\nu}}.e^{-i\pi\nu}
\label{8}
\end{equation}
where
\begin{equation}
R=\frac{M^{(1)}_{-\nu}(0,h^2)}{M^{(1)}_{\nu}(0,h^2)}
\label{9}
\end{equation}
This $S$--matrix describes the scattering of an incident wave (component
of the scalar field) of energy $\omega$ off the spherically
symmetric potential.
One could visualise this scattering as a spacetime
curvature effect or -- with black hole event horizon zero --
as that of a potential barrier surrounding the horizon.
With the horizon shrunk to zero at the origin (implying
in the field theory a relation between mass and charge reminiscent of
the Bogomol'nyi equation), this extremal case
corresponds to that of a BPS state.

The absorptivity is $A=1-SS^{\star}$.  The absorption cross section
differs from this by a multiplicative factor in front. The absorption
cross section $\sigma^l_{abs}$ of the $l$--th partial wave
in $n$ spatial dimensions has been derived in ref.\cite{12} and is given by
\begin{equation}
\sigma^l_{abs}=\frac{2^{n-2}\pi^{n/2-1}}{{\omega}^{n-1}}(n/2-2)!(l+n/2-1)
{l+n-3\choose l}(1-|S|^2)
\label{10}
\end{equation}
For $n=6$ as in our case this $l$--wave (here semiclassical)
absorption cross section (or socalled greybody factor) is
given by
\begin{equation}
\sigma^l_{abs}=\frac{8\pi^2}{3{\omega}^5}(l+1)(l+2)^2(1-|S|^2)
\label{11}
\end{equation}

\section
{The $D3$--brane in Born--Infeld theory}
\label{sec:III}

To supplement
the previous section, we consider briefly the
simplest version of supersymmetric Born--Infeld
electrodynamics for the $3$--brane. Our main
intention is to recall that the equation of small
fluctuations about the $D3$--brane
is again a modified Mathieu equation
as obtained above. In the simplest such model
reduced to the static case we write
the Lagrangian
\begin{eqnarray}
L=\int d^px{\cal L}, \;
{\cal L} = 1-\bigg[1-(\partial_i\phi)^2+(\partial_iy)^2+\nonumber\\
 (\partial_i\phi.\partial_iy)^2-(\partial_i\phi)^2
(\partial_jy)^2\bigg]^{\frac{1}{2}}
-\Sigma_pe\phi\delta({\bf r})
\label{d1}
\end{eqnarray}
Here $E_i=F_{0i}=-\partial_i\phi, i=1,...,p$ and $y(x_i)$
originates from one of the gauge field components
$A_a$ for $ a = p+1,...,d-1, d$= dimension, which
represent the transverse displacements of
the brane (of which we consider only one, e.g. $A_9$). The
source term of the electric field with charge $e$
and $\Sigma_3=4\pi$ hints at spherical symmetry.  Considering
only this case here and hence that of $S$--waves, we obtain
two Euler--Lagrange equations which we can write
$$
\partial_r\left(r^{p-1}\frac{\partial{\cal L}}{\partial(\partial_r y)}\right) = 0 , \;\;
r^{p-1}\frac{\partial{\cal L}}{\partial(\partial_r y)} = c
$$
where $c$ is a constant of integration.  Explicitly,
\begin{equation}
\frac {\phi^{\prime}}{\left[1-(\phi^{\prime})^2
+(y^{\prime})^2\right]^{\frac{1}{2}}}
=-\frac{e}{r^{p-1}},\;\;\;
\frac{-y^{\prime}}{\left[1-(\phi^{\prime})^2
+(y^{\prime})^2\right]^{\frac{1}{2}}}
=\frac{c}{r^{p-1}}
\label{d2}
\end{equation}
so that
\begin{equation} 
\frac{\phi^{\prime}}{y^{\prime}}=\frac{e}{c} \equiv \frac{1}{a}
\label{d3}
\end{equation}
Then
\begin{equation}
(\phi^{\prime})^2 = \frac{e^2}{r^{2(p-1)}+e^2(1-a^2)},\;\;\;
(y^{\prime})^2=\frac{(ea)^2}{r^{2(p-1)}+e^2(1-a^2)}
\label{d4}
\end{equation}
The $p$--brane and anti-$p$--branes are now given by
\begin{equation}
y(r)=\stackrel{+}{(-)}ae\int^{\infty}_rdr\frac{1}
{\sqrt{r^{2(p-1)}-r^{2(p-1)}_0}}
\label{d5}
\end{equation}
where $r^{2(p-1)}_0 = e^2(a^2-1)\geq 1$.
In view of the proportionality (\ref{d3}) the Lagrangian can be written 
\begin{equation}
{\cal L} = 1-\sqrt{1-(1-a^2)(\partial_i\phi)^2} -\Sigma_pe\phi \delta({\bf r})
\label{d6}
\end{equation}
The contribution to the energy not including the source term
is for $p=3$
$$E_{ns} = \int d^3x\left\{\frac{1}{\sqrt{1-(1-a^2)(\partial_i\phi)^2}}
 - 1\right\}
$$
Only for a charge $e$ which is kept fixed under a scale transformation is the
energy minimal in the limit $a^2\rightarrow 1$.  This
is the limit of the Bogomol'nyi bound
and hence for this value of $a^2$ the Born--Infeld
configuration, i.e. the $Dp$--brane or string is
classically stable, i.e. a nontopological BPS state.
The reason is that for $a=1$ we have $\phi^{\prime}=y^{\prime}$
which in the original context with $y=A_9$ implies
$F_{0r}=\partial_rA_9$.
This again has implications for the supersymmetry
variation of the gaugino $\chi$, the susy partner
of the gauge field, which is
$$
\delta\chi = \Gamma^{\mu\nu}F_{\mu\nu}\epsilon
$$
($\mu, \nu$ being the original 10 dimensional
indices and $\Gamma^{\mu\nu}$ the appropriate
combination of 10 dimensional Dirac matrices).  In
the case of only the electric field and the one excitation
under consideration the gaugino variation is
\begin{eqnarray}
\delta\chi&=&-(\Gamma^{0r}\partial_rA_0+\Gamma^{9r}\partial_rA_9)\epsilon
\nonumber\\
&\stackrel{a=1}{=}&-(\Gamma^{0r}+\Gamma^{9r})\partial_rA_0\epsilon
\label{d6.1}
\end{eqnarray}
Thus, since $1+\Gamma^{0r}\Gamma^{9r}$ is
a projection operator it is precisely for $a=1$ that 
the variation $\delta\chi$ can be zero
for some nonzero $\epsilon$, thus preserving a fraction
(half) of the number of supersymmetries. 

The energy $E_{ns}$ is infinite, but
integrating from $r=\delta$ to infinity so that
$y=ae/\delta, dE_{ns}/dy$, the energy per unit length
of the string  is finite, i.e.  
$$
\frac{dE_{ns}}{dy} = \frac{1}{2}(1-a^2) \frac{4\pi e}{a} = const.
$$
As shown in \cite{4}, unless $a=1$ supersymmetry is
completely broken (i.e. the supersymmetry variation
of the gaugino would not be preserved).  In this
limit the throat radius $r_0$ becomes smaller and
smaller, and the brane pair
moves further and further apart.
If one considers small fluctuations $\xi$
orthogonal to the brane and the string,
one obtains the small fluctuation equation \cite{4}
\begin{equation}
\triangle_r\xi + \Omega^2\left[1+\frac{e^2(p-2)^2}{r^{2(p-1)}}\right]\xi = 0
\label{d7}
\end{equation}
where $\Omega^2 \geq 0$ for stability.
The radial part of these equations is with $\xi=r^{-\frac{(p-1)}{2}}\psi $ and
angular momentum $l$
$$
\frac{d^2\psi}{dr^2} +\left[\frac{1}{r^2}\left\{l(l+p-2)-\frac{(p-1)(p-3)}{4}\right\}
+\Omega^2\left(1+\frac{e^2(p-2)^2}{r^{2(p-1)}}\right)\right]\psi=0
$$
Thus for $p=3$ and $ x=\Omega r, \kappa=e\Omega^2$ and for $S$-waves since
the string cannot depend on the angular variables of the worldvolume
\begin{equation}
\left(\frac{d^2}{dx^2}+1+\frac{\kappa^2}{x^4}\right)\psi = 0
\label{d8}
\end{equation}
This equation is an $S$-wave radial Schr\"odinger equation
for an attractive singular potential $\propto
x^{-4}$ but depends only on the single coupling
parameter $\kappa$ with constant positive Schr\"odinger energy.

\section
{Calculation of the Floquet exponent
in singular and nonsingular cases}
\label{sec:IV}

The Floquet exponent $\nu$ enters the discussion of the Mathieu
equation in view of the Bloch wave property of the modified Mathieu
function $Me_{\nu}(z,h)$
\begin{equation}
Me_{\nu}(z+i\pi,h)=e^{i\nu\pi}Me_{\nu}(z,h)
\label{C.1}
\end{equation}
The Floquet exponent can be introduced in
 several ways. In ref.\cite{10} (p.107)
$\nu$ is introduced by the relation
\begin{equation}
\cos\pi\nu=y_I(\pi,\lambda,h^2)
\label{C.2}
\end{equation}
where $y_I(x)$ is a fundamental solution of the (periodic) Mathieu equation
satisfying the boundary conditions $y_I(0)=1,y^{\prime}_I(0)=1$
and $\lambda$ is the eigenvalue
which, of course, is not necessarily
an integer. With
a perturbation theory ansatz for $y_I(\pi,\lambda,h^2)$ around
$h^2=0$ the following expansion is then shown to result
(cf. ref.\cite{10},  p.124):
\begin{eqnarray}
\cos\pi\nu &=&\cos\pi\sqrt{\lambda}+h^4\frac{\pi\sin\pi\sqrt{\lambda}}{4\sqrt{\lambda}(\lambda-1)}+\nonumber\\
&+& h^8\bigg[\frac{15\lambda^2-35\lambda+8}{64(\lambda-1)^3(\lambda-4)\lambda
\sqrt{\lambda}}\pi\sin\pi\sqrt{\lambda}-
\frac{\pi^2\cos\pi\sqrt{\lambda}}{32\lambda(\lambda-1)^2}\bigg]+O(h^{12})
\label{C.3}
\end{eqnarray}
Alternatively one can apply directly perturbation theory to
a trivial periodic solution of the limit $h^2\rightarrow 0$ as shown
in ref.\cite{6}.  In this case the following expansion is obtained
\begin{eqnarray}
\lambda &=& \nu^2 +\frac{h^4}{2(\nu^2-1)}
+\frac{(5\nu^2+7)h^8}{32(\nu^2-1)^3(\nu^2-4)}\nonumber\\
&+&\frac{(9\nu^4+58\nu^2+29)h^{12}}{64(\nu^2-1)^5(\nu^2-4)(\nu^2-9)}+O(h^{16})
\label{C.4}
\end{eqnarray}
This series can be reversed to yield $\nu$, i.e.
\begin{eqnarray}
\nu^2&=&\lambda-\frac{h^4}{2(\lambda-1)}-\frac{(13\lambda-25)h^8}
{32(\lambda-1)^3(\lambda-4)}\nonumber\\
&-&\frac{(45\lambda^3-455\lambda^2+1291\lambda-1169)h^{12}}
{64(\lambda-1)^5(\lambda-4)^2(\lambda-9)}+O(h^{16})
\label{C.5}
\end{eqnarray}
and so
\begin{equation}
\nu=\sqrt\lambda+\frac{h^4}{4(1-\lambda)\sqrt\lambda}
-\frac{(8-35\lambda+15\lambda^2)h^8}{64(\lambda-4)(\lambda-1)^3\lambda\sqrt\lambda}+O(h^{12})
\label{C.6}
\end{equation}
One can easily check the agreement with the above expression for
$\cos\pi\nu$  
by evaluating $\cos\pi\nu$ with $\nu$
of this expansion.  

An obvious feature of all of these expansions is that they are singular
for integral values of $\lambda$.  This behaviour is wellknown.  It
means that these expansions are in these cases
really asymptotic expansions for
large values of $\nu$ or $\lambda$ and can be used in such cases.
However, for other values the expansions can also be convergent
for sufficiently small values of $h^2$ as is shown in ref.\cite{10}.  
For values of $\nu$ close to an integer or $\sqrt{\lambda}$ close
to an integer
one has to expand around these as is also
mentioned in ref.\cite{10} (pp.124-125). We demonstrate this in the case
of $\nu$ almost equal to $2$. Thus we set 
$$
\nu=2+\delta, \;\;\;\sqrt \lambda=2 +\epsilon
$$
so that 
\begin{equation}
\cos\pi\nu=\cos\pi(2+\delta)=\cos2\pi.\cos\pi\delta=1-\frac{\pi^2\delta^2}{2}
+\cdot\cdot\cdot
\label{C.7}
\end{equation}
and consider the limit $\epsilon\rightarrow 0$.
Expanding the cosine and sine expressions appearing in 
eq.(\ref{C.3}) about $\lambda=4$ we have
($\lambda-4\approx 4\epsilon$)
$$
\cos\sqrt\lambda\pi=\cos 2\pi+(\lambda -4)(-\sin\sqrt\lambda\pi)_{\lambda=4}.
\frac{\pi}{2\sqrt\lambda}+\cdot\cdot\cdot = 1-\frac{(\lambda-4)^2\pi^2}{8}
+\cdot\cdot\cdot
$$
and
$$
\sin\sqrt{\lambda}\pi=\sin 2\pi+(\lambda-4)(\cos\sqrt\lambda\pi)_{\lambda =4}
.\frac{\pi}{2\sqrt\lambda}+\cdot\cdot\cdot=\frac{\pi(\lambda-4)}{2\sqrt\lambda}
+\cdot\cdot\cdot
$$
Substitution into eq.(\ref{C.3})
and considering the approach $\lambda\rightarrow 4$ gives
\begin{eqnarray}
\cos\pi\nu &=& \bigg(1-\frac{\pi^2}{8}(\lambda-4)^2+\cdot\cdot\cdot\bigg)
+\frac{h^4\pi^2(\lambda-4)}{4\sqrt\lambda(\lambda-1).2\sqrt\lambda}\nonumber\\
&+&h^8\bigg(\frac{(15\lambda^2-35\lambda+8)\pi^2(\lambda-4)}
{64(\lambda-1)^3(\lambda-4)\lambda\sqrt\lambda.2\sqrt\lambda}
-\frac{\pi^2.1}{32\lambda(\lambda-1)^2}\bigg)
+\cdot\cdot\cdot\nonumber\\
&=&\bigg(1-\frac{(\lambda-4)^2\pi^2}{8}+\cdot\cdot\cdot\bigg)
+\frac{h^4\pi^2}{8\lambda(\lambda-1)}\bigg((\lambda-4)+\cdot\cdot\cdot
\bigg)\nonumber\\
&+&\frac{h^8\pi^2}{128\lambda^2(\lambda-1)^3}\bigg((11\lambda^2-31\lambda+8)
+\cdot\cdot\cdot\bigg)+O(h^{12})
\label{C.8}
\end{eqnarray}
Hence (observe the cancellation of factors $(\lambda-4)$ in the term
of $O(h^8)$)in the limit $\epsilon\rightarrow 0$
\begin{equation} 
\cos\pi\nu=1+\frac{h^8\pi^2(11\times 16-31\times 4 +8)}
{2^74^23^3}+\cdot\cdot\cdot = 1+\frac{5\pi^2 h^8}{2^93^2}+\cdot\cdot\cdot 
\label{C.9}
\end{equation} 
Comparing the expansions (\ref{C.7}) and (\ref{C.9}) we obtain
$$
\delta = \pm i\frac{\sqrt{5}}{3}(h/2)^4
$$
We see that although the coefficients
of higher order terms of eq.(\ref{C.3})
contain factors $(\lambda - 4)$ in the denominators 
and so suggest divergences, the trigonometric factors
$\sin\pi\sqrt\lambda $ in the numerator always cancel these out
and thus yield a regular expansion for $\nu$ which is even
convergent within a certain domain around $h^2=0$.
One thus obtains the expansion
\begin{equation}
\nu = 2 -\frac{i\sqrt{5}}{3}(h/2)^4+\frac{7i}{108\sqrt{5}}(h/2)^8
+\frac{11851i}{31104\sqrt{5}}(h/2)^{12}\cdot\cdot\cdot
\label{C.10}
\end{equation}
This expansion has also been given in ref.\cite{1}.
Expansions around other integral values of $\sqrt{\lambda}$ are
obtained similarly. Of course, the higher the value
of this integer, the more terms at the
beginning of the series are identical with those
obtainable from the perturbation series 
(\ref{C.6}) above.
Thus in the case of $\sqrt\lambda=3, 4$, we obtain from the first two
terms of expansion (\ref{C.6})
$$
\nu = 3+\frac{h^4}{4(-8)3}+\cdot\cdot\cdot = 3 - \frac{(h/2)^4}{6}+
\cdot\cdot\cdot
$$
and similarly
$$
\nu = 4 -\frac{(h/2)^4}{15}+\cdot\cdot\cdot
$$
in agreement also with results of ref.\cite{1} up to
the given order, i.e.
\begin{eqnarray}
\sqrt\lambda=3:\;\;\;\nu=3-\frac{1}{6}(h/2)^4
+\frac{133}{4320}(h/2)^8
+\frac{311}{1555200}(h/2)^{12}+\cdot\cdot\cdot,\nonumber\\
\sqrt\lambda=4:\;\;\;\nu=4-\frac{1}{15}(h/2)^4-\frac{137}{27000}(h/2)^8
+\frac{305843}{680400000}(h/2)^{12}\cdot\cdot\cdot
\label{C.11}
\end{eqnarray}

\section
 {Relation between Dougall and standard Mathieu coefficients}
\label{sec:V}

One may wonder how the Mathieu function coefficients of Dougall
\cite{3}
which are used in refs.\cite{1,2}, are related to those
in modern standard literature such as ref.\cite{10}.
We therefore demonstrate their
precise connection here. It is crucial thereby
to distinguish between nonsingular or asymptotic cases
and singular cases, as we shall see.  We
begin with the nonsingular case and
calculate a coefficient given
in ref.\cite{10} (up to the first nonleading contribution) by starting from
Dougall's definition of his coefficients. We shall see that
the coefficients given in ref.\cite{10} obtained from simple
continued fraction solution of the basic recurrence relation
of the coefficients are in this case not only easier to derive
but have also a simpler form than the coefficients of Dougall.

The modified Mathieu function in terms of exponentials
is defined in ref.\cite{10} as the following sum
\begin{equation}
Me_{\nu}(z,h^2):=\sum^{\infty}_{r=-\infty}c^{\nu}_{2r}(h^2)e^{(\nu+2r)z}
\label{a.1}
\end{equation}
where $\nu\neq\pm 1, \pm 2, \cdot \cdot \cdot $.
In ref.\cite{10}(p. 131)
the following relation of general validity is given and used	
\begin{equation}
Me_{-\nu}(z,h)=Me_{\nu}(-z,h)
\label{a.2}
\end{equation}
This relation implies that
\begin{equation}
c^{\nu}_{2r}(h^2)=c^{-\nu}_{-2r}(h^2)\;\;\; and \;\;\;
\frac{c^{\nu}_{2r}(h^2)}{c^{\nu}_{0}(h^2)}=\frac{c^{-\nu}_{-2r}(h^2)}
{c^{-\nu}_{0}(h^2)}
\label{a.3}
\end{equation}
We shall see explicitly that this relation holds also in
what we call the singular case below.
Dougall\cite{3} defines in his work the solution corresponding
to $ Me_{\nu}(z,h)$ as
\begin{equation}
\frac{J(\nu, z)}{\phi(\nu/2)} = \sum^{\infty}_{r=-\infty}
(-1)^r\frac{\phi (r+\nu/2)}{\phi(\nu/2)}
e^{(\nu+2r)z}
\label{a.4}
\end{equation}
We therefore expect the equivalences
\begin{equation}
\frac{J(\nu,z)}{\phi(\nu/2)}=Me_{\nu}(z,h^2),\;\;\;\;
(-1)^n\frac{\phi(n+\nu/2)}{\phi(\nu/2)}=\frac{c^{\nu}_{2n}(h^2)}{c^{\nu}_{0}(h^2)}
\label{a.5}
\end{equation}

We now verify the latter of these relations for the case $n=1$
in the nonsingular case (i.e. for $\nu\neq$ integer$ +O(h^2)$), i.e. we
show that
\begin{equation}
-\;\;\frac{\phi(\nu/2+1)}{\phi(\nu/2)}\rightarrow \frac{c^{\nu}_2(h^2)}{c^{\nu}_0(h^2)}=
\frac{h^2}{4(\nu+1)}+\frac{(\nu^2+4\nu+7)h^6}{128(\nu+1)^3(\nu+2)(\nu-1)}+\cdot\cdot\cdot
\label{a.6}
\end{equation}
where the expression on the right is given in ref.\cite{10}(p.121). We also show thereby
that in leading order for small $h^2$
\begin{equation}
\frac{\phi(n+\nu/2)}{\phi(\nu/2)}=\frac{(h/2)^{2n} \nu !}{(n+\nu)!n!}\left(1+O(h^4)\right)
\label{a.7}
\end{equation}
in agreement with ref.\cite{10} (p.121).
The demonstration of agreement requires eqs. (\ref{C.4}), (\ref{C.5})
(cf. also \cite{10},p. 119), i.e.
\begin{equation}
s=l+2,\;\;\;\;s^2 = \nu^2+\frac{h^4}{2(\nu^2-1)}+\frac{(5\nu^2+7)h^8}{32(\nu^2-1)^3(\nu^2-4)}+\cdot\cdot\cdot
\label{a.8}
\end{equation}
or
\begin{equation}
\pm s = \nu+\frac{h^4}{4\nu(\nu^2-1)}+\cdot\cdot\cdot
\label{a.9}
\end{equation}
This general relation is an asymptotic expansion
 in $\nu$ (i.e. for $\nu$ large), and can be 
obtained perturbatively \cite{6}. It is
crucial, of course, to deal separately with values of $\nu$ close
to a singular value like $\nu = 2$ (see below).

Dougall defines his coefficients $\phi$ by an expansion, of which the
leading and next--to--leading contributions are\cite{3}
\begin{eqnarray}
&&\phi(n+\nu/2)=\frac{(h/2)^{2n+\nu}}
{(n+\nu/2+s/2)!(n+\nu/2-s/2)!}\bigg\{1-\nonumber\\
&&\sum^{\infty}_{p_1=0}
\frac{(h/2)^4}
{(n+\nu/2+s/2+1+p_1)}\nonumber\\
&&\frac{1}
{(n+\nu/2+s/2+2+p_1)(n+\nu/2-s/2+1+p_1)(n+\nu/2-s/2+2+p_1)}\nonumber\\
&&+\cdot\cdot\cdot\bigg\}\nonumber\\
\label{a.10}
\end{eqnarray}
Taking into account only the leading contribution, we have
\begin{equation}
\frac{\phi(n+\nu/2)}{\phi(\nu/2)}=\frac{(h/2)^{2n}(\nu/2+s/2))!(\nu/2-s/2)!}
{(n+\nu/2+s/2)!(n+\nu/2-s/2)!}
\label{a.11}
\end{equation}
Using
\begin{equation}
(-z)!=\frac{\pi}{(z-1)!\sin\pi z}
\label{a.12}
\end{equation}
and the approximation
$ s\approx \nu$ (cf. eq.(\ref{a.9})) we obtain
\begin{equation}
\frac{\phi(n+\nu/2)}{\phi(\nu/2)}=\frac{(h/2)^{2n}\nu !}{(n+\nu)!n!}\bigg(1
+O(h^4)\bigg)
\label{a.13}
\end{equation} 
in agreement with $c^{\nu}_{2n}(h^2)/{c^{\nu}_0(h^2)}$ of ref.
\cite{10}. We see therefore
that the expansion (\ref{a.9}) plays an important role in establishing the
connection between the coefficients $\phi(r+\nu/2)$ and $c^{\nu}_{2r}$
in the nonsingular case.

We now consider the next--to--leading contribution in eq.(\ref{a.10}).
Setting
\begin{equation}
A^{(1)}_{q}= \sum^{\infty}_{p_1=0}\frac{1}
{(q+s/2+1+p_1)
(q+s/2+2+p_1)(q-s/2+1+p_1)(q-s/2+2+p_1)}
\label{a.14}
\end{equation}
we can make the partial fraction separation
\begin{eqnarray}
A^{(1)}_{q}&=& \sum^{\infty}_{p_1=0}\bigg[\frac{1}{s(s-1)}\left(\frac{1}{p_1+q+1+s/2}-\frac{1}{p_1+q+2-s/2}\right)\nonumber\\
&-&\frac{1}{s(s+1)}\left(\frac{1}{p_1+q+2+s/2}-\frac{1}{p_1+q+1-s/2}\right)\bigg]
\label{a.15}
\end{eqnarray}
The sums over individual terms are divergent.  Thus the expression
depends crucially on taking differences.  In order to deal with these
we use the formula \cite{13}
\begin{equation}
\sum^{n-1}_{k=0}\frac{1}{k+y}=\psi(n+y)-\psi(y)
\label{a.16}
\end{equation}
where $\psi (y)$ is the derivative of the log of the gamma function
$\Gamma (y)$. Thus
$$
\sum^{n-1}_{p_1=0}\left(\frac{1}{p_1+q+1+s/2}-\frac{1}{p_1+q+2-s/2}\right)
\stackrel{n\rightarrow\infty}{\longrightarrow}\psi(q+2-s/2) - \psi(q+1+s/2)
$$
and
$$
 \sum^{n-1}_{p_1=0}
\left(\frac{1}{p_1+q+2+s/2}-\frac{1}{p_1+q+1-s/2}\right)
\stackrel{n\rightarrow\infty}{\longrightarrow}\psi(q+1-s/2) - \psi(q+2+s/2)
$$
so that
\begin{eqnarray}
A^{(1)}_{q}&=&\frac{1}{s(s-1)}\bigg(\psi(q+2-s/2)-\psi(q+1+s/2)\bigg)\nonumber\\
&-&\frac{1}{s(s+1)}\bigg(\psi(q+1-s/2)-\psi(q+2+s/2)\bigg)
\label{a.17}
\end{eqnarray}
We now use eq.(\ref{a.9}) in order to reexpress $s$ interms of $\nu$.  Then
\begin{eqnarray}
\frac{1}{s(s-1)}&=&\frac{1}
{\nu(\nu-1)}\bigg[1-\frac{(2\nu-1)h^4}{4\nu^2(\nu-1)^2(\nu+1)}+\cdot\cdot\cdot\bigg]\nonumber\\
\frac{1}{s(s+1)}&=&\frac{1}
{\nu(\nu+1)}\bigg[1-\frac{(2\nu+1)h^4}{4\nu^2(\nu-1)(\nu+1)^2}
+\cdot\cdot\cdot\bigg]
\label{a.18}
\end{eqnarray}
Setting $q=n+\nu/2$ and dealing similarly with the arguments
of the functions $\psi$, we obtain in lowest order of $h$
\begin{eqnarray}
A^{(1)}_{q=n+\nu/2}&=&\frac{1}{\nu(\nu-1)}\bigg[\psi(n+2)-\psi(\nu+n+1)\bigg]
\nonumber\\
&-&\frac{1}{\nu(\nu+1)}\bigg[\psi(n+1)-\psi(\nu+n+2)\bigg]
\label{a.19}
\end{eqnarray}
Again we consider a difference, i.e. 
$$
\triangle_n:=A^{(1)}_{n+\nu/2}-A^{(1)}_{n-1+\nu/2} 
$$
We now use the formula
\begin {equation}
\psi(n+1)=-C+1+\frac{1}{2}+\frac{1}{3}+\cdot\cdot\cdot +\frac{1}{n}
\label{a.20}
\end{equation}
where $C$ is the Euler constant.  Then
$$
\psi(n+2)-\psi(n+1)=\frac{1}{n+1},\;\;\;\; \psi(n+1)-\psi(n)=\frac{1}{n}
$$
and $\triangle_n$ becomes
\begin{eqnarray}
\triangle_n&=&\frac{1}
{\nu(\nu-1)}\bigg[\frac{1}{n+1}-\psi(\nu+n+1)+\psi(\nu+n]\bigg]
\nonumber\\
&-&\frac{1}{\nu(\nu+1)}\bigg[\frac{1}{n}-\psi(\nu+n+2)+\psi(\nu+n+1)\bigg]
\label{a.21}
\end{eqnarray}
We now require yet another formula of the $\psi$ function, i.e.
\begin{equation}
\psi(x)= - C +\sum^{\infty}_{i=0}\left(\frac{1}{n+1}-\frac{1}{x+n}\right)
\label{a.22}
\end{equation}
With this we obtain
$$
\psi(\nu+n)-\psi(\nu+n+1)=-\frac{1}{\nu+n},\;\;\;
\psi(\nu+n+1)-\psi(\nu+n+2)=-\frac{1}{\nu+n+1}
$$
(where in each case the dummy summation index of the second sum, i.e. $i$,
was renamed $i-1$).
We therefore obtain in the dominant approximation
\begin{equation}
\triangle_n = \frac{1}{\nu(\nu-1)}\bigg[\frac{1}{n+1}-\frac{1}{\nu+n}\bigg]
-\frac{1}{\nu(\nu+1)}\bigg[\frac{1}{n}-\frac{1}{\nu+n+1}\bigg]
\label{a.23}
\end{equation}
For $n=1$ this implies
\begin{equation}
\triangle_1=A^{(1)}_{q=1+\nu/2}-A^{(1)}_{q=\nu/2}= -\frac{1}{2(\nu+1)(\nu+2)}
\label{a.24}
\end{equation}
in the dominant approximation.
This difference will now have to be substituted into the Dougall
coefficient
\begin{eqnarray}
\frac{\phi(\nu/2+1)}{\phi(\nu/2)}&=&
\frac{(\frac{h}{2})^2(\nu/2+s/2)!(\nu/2-s/2)!
\bigg\{1-
(\frac{h}{2})^4A^{(1)}_{\nu/2+1}+\cdot\cdot\cdot\bigg\}}{(\nu/2+1+s/2)!
(\nu/2+1-s/2)!\bigg\{1-(\frac{h}{2})^4A^{(1)}_{\nu/2}+\cdot\cdot\cdot\bigg\}}\nonumber\\
&\approx&{\left(\frac{h}{2}\right)}^2
\frac{\bigg\{1-{\left(\frac{h}{2}\right)^4}
 \bigg(A^{(1)}_{\nu/2+1}-A^{(1)}_{\nu/2}\bigg)+\cdot\cdot\cdot\bigg\}}
{(\nu/2+1+s/2)(\nu/2+1-s/2)}
\label{a.25}
\end{eqnarray}
Inserting eq.(\ref{a.24}) and for $s$ the series of eq.(\ref{a.9})
we obtain
\begin{equation}
\frac{\phi(\nu/2+1)}{\phi(\nu/2)}=\frac{h^2}{4(\nu+1)}+\frac{h^6(\nu^2+4\nu+7)}{128(\nu-1)(\nu+1)^3(\nu+2)}+O(h^{10})
=-\frac{c^{\nu}_2}{c^{\nu}_0}
\label{a.26}
\end{equation}
in agreement with ref.\cite{10}(p.121). To obtain the Dougall coefficients in this
form is thus seen to
be rather complicated. This may explain why Dougall himself does not evaluate
any of his coefficients explicitly.

In the remainder of this section we calculate with the method of Dougall
the important coefficients $\phi(\pm \nu/2)$ for
the $S$--wave case, i.e. $s=2$, and demonstrate the agreement
with results of ref. \cite{1}.  The results will also
be needed in the next section in establishing ratios
corresponding to the ratio of eq.(\ref{a.26}). 
From eq.(\ref{a.17}) we obtain
for the leading term in the
limit $h^2\rightarrow 0$ for $q = \nu/2$ and $s = 2$ (i.e. $\nu\approx 2$,
a socalled singular case)
\begin{eqnarray}
A^{(1)}_{\nu/2}&=&\bigg\{\frac{1}{2}\bigg[\psi(\nu/2+1)-\psi(\nu/2+2)\bigg]
-\frac{1}{6}\bigg[\psi(\nu/2)-\psi(\nu/2+3)\bigg]\bigg\}_{h^2\rightarrow 0}  
\nonumber\\
&=&\frac{1}{2}\bigg[\psi(2)-\psi(3)\bigg]-\frac{1}{6}\bigg[\psi(1)-\psi(4)\bigg]
\label{a.27}
\end{eqnarray}
Using eq.(\ref{a.20}) we obtain
$$
A^{(1)}_{\nu/2}=\frac{1}{18}
$$
The authors of ref.\cite{1} developed another algorithm in which (cf. their
Appendix A)
\begin{equation}
A^{(1)} = A^{(1)}_z \equiv S[1]
\label{a.28}
\end{equation}
and for $s=2$ (their $r=1$)
\begin{equation}
A^{(1)}_z = \frac{3+2z}{3z(2+z)}+\frac{\psi(z)-\psi(z+2)}{3}
\label{a.29}
\end{equation}
Here
\begin{equation}
\psi(z)-\psi(z+2)=\frac{d}{dz}\bigg[\ln\Gamma (z)-\ln\Gamma (z+2)\bigg]
=-\frac{2z+1}{z(z+1)}
\label{a.30}
\end{equation}
so that
\begin{equation}
A^{(1)}_z=\frac{1}{3z(z+1)(z+2)}
\label{a.31}
\end{equation}
and for $\nu\approx 2$ one obtains again 
$$
A^{(1)}_{\nu/2}= \frac{1}{18}
$$
as above.
It follows that for this case with $\Re \nu >0$ (cf. eq.(\ref{a.10}),
\begin{eqnarray}
\phi(\nu/2)&\approx &\frac{(h/2)^{\nu}}{(\nu/2+s/2)!(\nu/2-s/2)!}\bigg\{1
-\frac{(h/2)^4}{18}\bigg\}\nonumber\\
&\approx &\frac{1}{2}(h/2)^2\bigg[1+O(h^4)\bigg]
\label{a.32}
\end{eqnarray}
Fortunately for this case the expansion (\ref{a.26}) does not seem to
possess terms which diverge for $\nu$ close to a positive
integer$\neq 1$, and evidently even if it did -- since in our cases
$\nu =$ an integer $\neq 1 +O(h^4)$ -- would not affect the leading term of
eq.(\ref{a.28}).  This is radically different
when $\nu$ is close to a negative integer such as $-2$ in that case.
We see from (\ref{a.22}) that if $x= -2 +O(h^4)$, $\psi\propto
1/h^4$, and hence $A^{(1)}_q \propto 1/h^4$, and so this term
will contribute to the leading factor in the coefficient $\phi$
of eq.(\ref{a.10}). 

We now consider the coefficient $\phi(-\nu/2), \Re\nu>0$, for
which
\begin{equation}
s=2,\;\;\;
\nu=2-\frac{i\sqrt{5}}{3}(h/2)^4+\cdot\cdot\cdot
\label{a.33}
\end{equation}
and calculate this first with the
method of Dougall.
The following steps given
explicitly demonstrate clearly
how singularities in the limit
$h^2\rightarrow 0$ arise and how they have to
be handled.
Thus with eq.(\ref{a.10}):
\begin{eqnarray}
\phi(-\nu/2)&=&\frac{(h/2)^{-\nu}}
{(-\nu/2+s/2)!(-\nu/2-s/2)!}\bigg\{1\nonumber\\
&-&\sum^{\infty}_{p_1=0}\frac{(h/2)^4}
{(-\nu/2+s/2+1+p_1)(-\nu/2+s/2+2+p_1)}\nonumber\\
&\cdot &\frac{1}{(-\nu/2-s/2+1+p_1)(-\nu/2-s/2+2+p_1)}+\cdot\cdot\cdot\bigg\}
\nonumber\\
&\approx &\frac{(h/2)^{-2}}{0!(-2+\frac{i\sqrt{5}}{6}(h/2)^4)!}\bigg\{1
-(h/2)^4\cdot
\nonumber\\
&\cdot\bigg [&\frac{1}
{(1)(2)(-1+\frac{i\sqrt{5}}{6}(h/2)^4)
(\frac{i\sqrt{5}}{6}(h/2)^4)}\nonumber\\
&+&\frac{1}{(2)(3)(\frac{i\sqrt{5}}{6}(h/2)^4)
(1+\frac{i\sqrt{5}}{6}(h/2)^4)}\bigg ]\nonumber\\
&+&\sum^{\infty}_{p_1=2}\cdot\cdot\cdot \bigg\}\nonumber\\
&\approx &\frac{(h/2)^{-2}}{(-2+\frac{i\sqrt{5}}{6}(h/2)^4)!}\bigg\{1
-(h/2)^4\bigg[\frac{1}{(-2)(\frac{i\sqrt{5}}{6}(h/2)^4)}+\frac{1}{6(\frac{i\sqrt{5}}{6}(h/2)^4)}\bigg]+\cdot\cdot\cdot\bigg\}\nonumber\\
&=&\frac{(h/2)^{-2}}{(-2+\frac{i\sqrt{5}}{6}(h/2)^4)!}\bigg\{1+\frac{2}{i\sqrt{5}}\bigg\}\nonumber\\
&\stackrel{eq.(\ref{a.12})}{=}&\frac{1}{\pi}(h/2)^{-2}\bigg\{1+\frac{2}{i\sqrt{5}}\bigg\}(1)(-1)\pi \frac{i\sqrt{5}}{6}(h/2)^4\nonumber\\
&=&-(h/2)^2\bigg(\frac{2+i\sqrt{5}}{6}\bigg)
\label{a.34}
\end{eqnarray}

The steps above clearly show how the singular terms in the next--to--leading
contribution contribute to the dominant order.

The result 
(\ref{a.34})
will now be shown to agree with the calculations of the method of ref.\cite{1}. 
For this purpose we set in eq.(\ref{a.31}) $z=-\nu/2$ and replace $\nu$
by the expression in eq.(\ref{a.33}).  Then
$$
A^{(1)}_{-\nu/2}\approx\frac{-8}{3.2(i\sqrt{5}/3).2.(h/2)^4}=-\frac{2}{i\sqrt{5}(h/2)^4}
$$
and 
$$
v\equiv 1-(h/2)^4A^{(1)}_{-\nu/2}=\frac{2+i\sqrt{5}}{i\sqrt{5}}
$$
From eq.(\ref{a.10})we obtain therefore
\begin{equation}
\phi(-\nu/2)=\frac{(h/2)^{-2}}{0!(-2+\frac{i\sqrt{5}}{6}(h/2)^4)!}.v
\label{a.35}
\end{equation}
Using for the factorial again eq.(\ref{a.12}), one obtains
\begin{equation}
\phi(-\nu/2)=-\frac{(h/2)^{-2}\frac{i\sqrt{5}}{6}(h/2)^4\pi}{\pi}.\bigg\{1+
\frac{2}{i\sqrt{5}}\bigg\}
\label{a.36}
\end{equation}
Thus finally
\begin{equation}
\phi(-\nu/2)=-\frac{2+i\sqrt{5}}{6}(h/2)^2
\label{a.37}
\end{equation}
in agreement with eq.(\ref{a.34}).

\section{ Calculation of standard Mathieu coefficients}
\label{sec:VI}

Our next objective is to compute a Dougall coefficient
(i.e. a ratio of two quantities $\phi$) and to
compare it with a standard Mathieu coefficient in the nontrivial
singular case.  As a suitable example we choose
the coefficient $\phi(-\nu/2+1)$ which when divided by $\phi(-\nu/2)$
(calculated above) ought to agree with the Mathieu coefficient
$-c^{-\nu}_2/c^{-\nu}_0$
according to eq.(\ref{a.5}), i.e. we wish to demonstrate that
\begin{equation}
\phi(-\nu/2+1)/\phi(-\nu/2)=-c^{-\nu}_2/c^{-\nu}_0
\label{a.38}
\end{equation}
We begin with the calculation of the Dougall coefficient. Using
eq.(\ref{a.10}), we have
\begin{eqnarray}
\phi(-\nu/2+1)&=&\frac{(h/2)^{2-\nu}}{(1-\nu/2+s/2)!(1-\nu/2-s/2)!}\bigg\{1
\nonumber\\
&-&\sum^{\infty}_{p_1=0}\frac{(h/2)^4}{(1-\nu/2+s/2+1+p_1)
(1-\nu/2+s/2+2+p_1)}\nonumber\\
&\cdot&\frac{1}{(1-\nu/2-s/2+1+p_1)(1-\nu/2-s/2+2+p_1)}+\cdot\cdot\cdot
\bigg\}\nonumber\\
&\stackrel{eq.(\ref{a.33}) for  \nu}{=}&\frac{1}{(1)(-1+\frac{i\sqrt{5}}{6}(h/2)^4)!}\bigg\{1-(h/2)^4\bigg[\frac{1}{(2)(3)(
\frac{i\sqrt{5}}{6}(h/2)^4)(1)}\nonumber\\
&+&\frac{1}{(3)(4)(1)(2)}\bigg]+\sum^{\infty}_{p_1=2}\cdot\cdot\cdot\bigg\}\nonumber\\
&\approx&\frac{(0)!\sin\pi(1-\frac{i\sqrt{5}}{6}(h/2)^4)}{\pi}\bigg\{1
-\frac{1}{i\sqrt{5}}\bigg\}\nonumber\\
&\approx&\frac{i\sqrt{5}-1}{6}(h/2)^4
\label{a.39}
\end{eqnarray}
With eq.(\ref{a.37}) we obtain therefore
\begin{equation}
\frac{\phi(-\nu/2+1)}{\phi(-\nu/2)}=(h/2)^2\bigg(\frac{1-i\sqrt{5}}{2+i\sqrt{5}}
\bigg)\bigg[1+O(h^4)\bigg]
\label{a.40}
\end{equation}
Similarly one obtains
\begin{equation}
\phi(\nu/2+1)=\frac{1}{6}(h/2)^4\bigg[1+O(h^4)\bigg],\;\;\;
\frac{\phi(\nu/2+1)}{\phi(\nu/2)}=\frac{1}{3}(h/2)^2\bigg[1+O(h^4)\bigg]
\label{a.41}
\end{equation}

Our next step is to derive the corresponding expression from the
continued fraction relation of the recurrence relation of
the standard Mathieu coefficients.  This recurrence relation is
given by (cf. ref.\cite{10}, p. 117)
\begin{equation}
\frac{c^{\nu}_{2r}}{c^{\nu}_{2r-2}}=\frac{1}
{\displaystyle h^{-2}[s^2-(\nu+2r)^2]-\frac{1}
{\displaystyle h^{-2}[s^2-(\nu+2r+2)^2]-\frac{1}
{\displaystyle h^{-2}[s^2-(\nu+2r+4)^2]\cdot\cdot\cdot}}}
\label{a.42}
\end{equation}
Here we set $r=1$ and replace $\nu$ by $-\nu$. Then we again
use (\ref{a.33}) and a) set  $s=2$, and b) replace $\nu$
by the expansion given in eq.(\ref{a.33}). 
One then has to go as far as
the terms explicitly written out
in the following continued fraction
only to obtain the dominant contribution:
\begin{equation}
\frac{c^{-\nu}_{2}}{c^{-\nu}_{0}}=\frac{1}{\displaystyle h^{-2}
[s^2-(-\nu+2)^2]-\frac{1}
{\displaystyle h^{-2}[s^2-(-\nu+4)^2]-\frac{1}
{\displaystyle h^{-2}[s^2-(-\nu+6)^2]\cdot\cdot\cdot}}}
\label{a.43}
\end{equation}
Making the substitutions we obtain
\begin{equation}
\frac{c^{-\nu}_{2}}{c^{-\nu}_{0}}=\frac{1}
{\displaystyle h^{-2}[4]-\frac{1}
{\displaystyle h^{-2}[-\frac{4i\sqrt{5}}
{3}(h/2)^4]-\frac{1}{\displaystyle h^{-2}[-12]\cdot\cdot\cdot}}}
\label{a.44}
\end{equation}
This can be seen to reduce to
\begin{equation}
\frac{c^{-\nu}_{2}}{c^{-\nu}_{0}}=(h/2)^2\frac{i\sqrt{5}-1}{i\sqrt{5}+2}
\label{a.45}
\end{equation}
which agrees with the negative of the above Dougall coefficient as expected
on the basis of eq.(\ref{a.4}). 
One can see that the calculation here is
simpler than that of both (\ref{a.37}) and (\ref{a.39}). 

The reciprocal of the continued fraction relation (\ref{a.42}) is
(cf. ref.\cite{10}, p.117)
\begin{equation}
\frac{c^{\nu}_{2r-2}}{c^{\nu}_{2r}}=\frac{1}{\displaystyle h^{-2}
[s^2-(\nu+2r-2)^2]-\frac{1}
{\displaystyle h^{-2}[s^2-(\nu+2r-4)^2]-\frac{1}
{\displaystyle h^{-2}[s^2-(\nu+2r-6)^2]\cdot\cdot\cdot}}}
\label{a.46}
\end{equation}
Replacing here $\nu$ by $-\nu$ we have
\begin{equation}
\frac{c^{-\nu}_{2r-2}}{c^{-\nu}_{2r}}=\frac{1}{\displaystyle h^{-2}
[s^2-(-\nu+2r-2)^2]
-\frac{1}
{\displaystyle h^{-2}[s^2-(-\nu+2r-4)^2]-\frac{1}
{\displaystyle h^{-2}[s^2-(-\nu+2r-6)^2]\cdot\cdot\cdot}}}
\label{a.47}
\end{equation} 
Here we put $r=0$ and again make the replacements of eq.(\ref{a.33}).
Then
\begin{equation}
\frac{c^{-\nu}_{-2}}{c^{-\nu}_{0}}=-\frac{h^2}{12}\bigg[1+O(h^4)\bigg]
\label{a.48}
\end{equation}
For $r=2$ in eq.(\ref{a.47}) we obtain
\begin{equation}
\frac{c^{-\nu}_{4}}{c^{-\nu}_{2}}=\frac{12}{h^2(1-i\sqrt{5})}
\label{a.49}
\end{equation}
so that with eq. (\ref{a.45})
\begin{equation}
\frac{c^{-\nu}_{4}}{c^{-\nu}_{0}}=\frac{c^{-\nu}_{4}}{c^{-\nu}_{2}}
\frac{c^{-\nu}_{2}}{c^{-\nu}_{0}}=-\frac{3}{2+i\sqrt{5}}
\label{a.50}
\end{equation}
In a similar way we obtain
\begin{equation}
\frac{c^{-\nu}_{-4}}{c^{-\nu}_{-2}}=-\frac{h^2}{2^5}\bigg[1+O(h^4)\bigg],
\;\;\;\frac{c^{-\nu}_{-4}}{c^{-\nu}_{0}}
=\frac{c^{-\nu}_{-4}}{c^{-\nu}_{-2}}
\frac{c^{-\nu}_{-2}}{c^{-\nu}_{0}}=\frac{h^4}{3.2^7}\bigg[1+O(h^4)\bigg]
\label{a.51}
\end{equation}

Summarising we have as leading contributions of
standard Mathieu coefficients in the singular case $l=0$ or $s=2$:
\begin{eqnarray}
\frac{c^{-\nu}_2}{c^{-\nu}_0}&=&(h/2)^2\frac{i\sqrt{5}-1}{i\sqrt{5}+2}
\bigg[1+O(h^4)\bigg]=\frac{c^{\nu}_{-2}}{c^{\nu}_0}\nonumber\\
\frac{c^{-\nu}_{-2}}{c^{-\nu}_0}&=&-\frac{(h/2)^2}{3}\bigg[1+O(h^4)\bigg]
=\frac{c^{\nu}_{2}}{c^{\nu}_0}\nonumber\\
\frac{c^{-\nu}_4}{c^{-\nu}_0}&=&-\frac{3}{2+i\sqrt{5}}\bigg[1+O(h^4)\bigg]
=\frac{c^{\nu}_{-4}}{c^{\nu}_0}\nonumber\\
\frac{c^{-\nu}_{-4}}{c^{-\nu}_0}&=&\frac{(h/2)^4}{2^3.3}\bigg[1+O(h^4)\bigg]
=\frac{c^{\nu}_{4}}{c^{\nu}_0}\nonumber\\
\frac{c^{-\nu}_{6}}{c^{-\nu}_0}&=&(h/2)^2\frac{1}{2+i\sqrt{5}}\bigg[1+O(h^4)\bigg]
=\frac{c^{\nu}_{-6}}{c^{\nu}_0}\nonumber\\
\frac{c^{-\nu}_{-6}}{c^{-\nu}_0}&=&-\frac{(h/2)^6}{2^3.3^2.5}
\bigg[1+O(h^4)\bigg]
=\frac{c^{\nu}_{6}}{c^{\nu}_0}
\label{a.52}
\end{eqnarray}
In Appendix B we give several more terms of these expansions calculated
with Mathematica.

\section{Evaluation of the quantity $R$ in the singular case}
\label{sec:VII}

Having determined the standard Mathieu coefficients in the singular
S--wave case, we can proceed to evaluate the quantity $R$ entering
the S--matrix.  $R$ was defined in ref.\cite{6}(see also
Appendix A) as
\begin{equation}
R=\alpha_{\nu}/\alpha_{-\nu},\;\;\;\; \alpha_{\nu}(h)=
Me_{\nu}(0,h)/M^{(1)}_{\nu}(0,h)
\label{B.1}
\end{equation}
The function $Me_{\nu}(z,h)$ was defined previously.  The functions
$M^{(i)}_{\nu}(z,h)$, for $i=1,2,3,4$, are corresponding expansions
of the modified Mathieu function in terms of cylindrical functions
$J_{\nu}(z),Y_{\nu}(z),H^{(1)}_{\nu}(z),H^{(2)}_{\nu}(z)$ respectively.
In particular we have the expansion (cf. ref.\cite{10}, p. 178)
\begin{equation}
Me_{\nu}(0,h)M^{(1)}_{\nu}(z,h)=\sum^{\infty}_{r=-\infty}c^{\nu}_{2r}(h^2)
J_{\nu+2r}(2h\cosh z)
\label{B.2}
\end{equation}
As shown in ref. \cite{10}(p. 180),
a much better expansion to use in practice for
$ M^{(1)}_{\nu}(z,h)$ in view
of its rapid convergence, is
\begin{equation}
c^{\pm\nu}_{2r}(h^2) M^{(1)}_{\pm\nu}(z,h)
=\sum^{+\infty}_{l=-\infty}(-1)^lc^{\pm\nu}_{2l}(h^2)
J_{l-r}(he^{-z})J_{\pm\nu+l+r}(he^{z})
\label{B.3}
\end{equation}
so that in particular
\begin{equation}
c^{\pm\nu}_{2r}(h^2) M^{(1)}_{\pm\nu}(0,h)
=\sum^{+\infty}_{l=-\infty}(-1)^lc^{\pm\nu}_{2l}(h^2)
J_{l-r}(h)J_{\pm\nu+l+r}(h)
\label{B.4}
\end{equation}
This formula is amazing.  It implies that one and the same
quantity $M^{(1)}_{\pm\nu}(0,h)$ can be obtained from many different
expansions (and so different Bessel functions)  by allocating
different values to $r$, i.e. e.g. $r=0$ and $2$. An analogous observation
has also been made by Dougall\cite{3}.  

We begin with the evaluation of $ Me_{\nu}(0,h)$, i.e.
\begin{equation}
 Me_{\nu}(0,h)=c^{\nu}_0(h^2)\sum_r\frac{c^{\nu}_{2r}(h^2)}{c^{\nu}_0(h^2)}
\label{B.5}
\end{equation}
With the help of the standard Mathieu coefficients evaluated previously
we obtain
\begin{equation}
\bigg [Me_{\nu}(0,h)\bigg]_{h^2\rightarrow 0}
=c^{\nu}_0(0)\bigg[1-\frac{3}{2+i\sqrt{5}}\bigg]
=c^{\nu}_0(0)\bigg[\frac{i\sqrt{5}-1}{2+i\sqrt{5}}\bigg]
=\bigg[Me_{-\nu}(0,h)\bigg]_{h^2\rightarrow 0}
\label{B.6}
\end{equation}
The last equality follows also from the general property
$ Me_{-\nu}(z,h)= Me_{\nu}(-z,h)$ (cf. ref.\cite{10}, p. 131).
More terms can be calculated with Mathematica.
Thus 
$$
Me_\nu (0,h)=1+\frac{c_2^\nu (h^2)}{c_0^\nu (h^2)}+\frac{c_4^\nu (h^2)}{%
c_0^\nu (h^2)}+\frac{c_6^\nu (h^2)}{c_0^\nu (h^2)}+\frac{c_{-2}^\nu (h^2)}{%
c_0^\nu (h^2)}+\frac{c_{-4}^\nu (h^2)}{c_0^\nu (h^2)}+\frac{c_{-6}^\nu (h^2)%
}{c_0^\nu (h^2)}+\frac{c_{-8}^\nu (h^2)}{c_0^\nu (h^2)}+\frac{c_{-10}^\nu
(h^2)}{c_0^\nu (h^2)} 
$$
We have to take into account also  $\frac{c_{-8}^\nu (h^2)}{c_0^\nu (h^2)}$
 and $\frac{%
c_{-10}^\nu (h^2)}{c_0^\nu (h^2)}$ because these contribute to orders
$h^4$ and $h^6$, i.e. 
\begin{eqnarray*}
\frac{c_{-8}^\nu (h^2)}{c_0^\nu (h^2)} &=&\frac{(i-\sqrt{5})}{2^33(i+\sqrt{5}%
)}\left( \frac h2\right) ^4+\frac{202i+35\sqrt{5}}{2^43^35(i+\sqrt{5})}%
\left( \frac h2\right) ^8 \\
\frac{c_{-10}^\nu (h^2)}{c_0^\nu (h^2)} &=&\frac{-i+\sqrt{5}}{2^33^25(i+%
\sqrt{5})}\left( \frac h2\right) ^6
\end{eqnarray*}
respectively.
We then obtain the following result correct up to order $h^6$: 
\begin{eqnarray}
Me_\nu (0,h)&=&\frac{2i}{i+\sqrt{5}}+\frac{4i}{3(i+\sqrt{5})}\left( \frac h2%
\right) ^2+\frac{65i+11\sqrt{5}}{2^33^15(i+\sqrt{5})}\left( \frac h2\right)
^4\nonumber\\
&+&\frac{11(83i+89\sqrt{5})}{2^23^35(19i-5\sqrt{5})}\left( \frac h2\right)
^6+O(h^7) 
\label{B.7}
\end{eqnarray}
We note here that for general values of $\nu $ not equal to an integer
one obtains (up to and including contributions of $O(h^6)$)
\begin{eqnarray}
Me_\nu (0,h)&=&1+\frac 2{\nu ^2-1}\left( \frac h2\right)
^2+\frac{\nu ^2+2}{%
(\nu ^2-1)(\nu ^2-4)}\left( \frac h2\right)
^4\nonumber\\
&+&\frac{2(\nu ^6+4\nu ^4-39\nu
^2-62)}{(\nu ^2-1)^3(\nu ^2-4)(\nu ^2-9)}\left( \frac
h2\right) ^6+O(h^7)
\label{B.71}
\end{eqnarray}
Next we evaluate  $M^{(1)}_{\nu}(0,h)$ with the help of  
eq.(\ref{B.4}) choosing $r=0$ and then as a check $r=2$. In the first
case we obtain the expansion
\begin{eqnarray}
M^{(1)}_{\nu}(0,h)&=& J_0(h)J_{\nu}(h)-\frac{c^{\nu}_2(h^2)}{c^{\nu}_0(h^2)}
J_1(h)J_{\nu+1}(h)
-\frac{c^{\nu}_{-2}(h^2)}{c^{\nu}_0(h^2)}J_{-1}(h)J_{\nu-1}(h)\nonumber\\
&+&\frac{c^{\nu}_4(h^2)}{c^{\nu}_0(h^2)}J_2(h)J_{\nu+2}(h)
+\frac{c^{\nu}_{-4}(h^2)}{c^{\nu}_0(h^2)}J_{-2}(h)J_{\nu-2}(h)+\cdot\cdot\cdot
\label{B.8}
\end{eqnarray}
In lowest orders of $h^2$ this is
$$
M^{(1)}_{\nu}(0,h)= J_0(h)J_{\nu}(h)
+\frac{c^{\nu}_{-4}(h^2)}{c^{\nu}_0(h^2)}J_{-2}(h)J_{\nu-2}(h)
$$
which when evaluated in lowest orders of $h^2$ implies
$$
\frac{(h/2)^2}{2}.1+\frac{-3}{2+i\sqrt{5}}.\frac{(h/2)^2}{2! 0!}.1
$$
It follows that
\begin{equation}
M^{(1)}_{\nu}(0,h)=\frac{1}{2}(h/2)^2
\bigg[\frac{-1+i\sqrt{5}}{2+i\sqrt{5}}\bigg] 
\label{B.9}
\end{equation}
Here, of course, $\nu$ has as before the S--wave value, i.e.
\begin{equation}
\nu = 2 -\frac{i\sqrt{5}}{3}(h/2)^4+\cdot\cdot\cdot
\label{B.10}
\end{equation}
If we set $r=2$ in eq.(\ref{B.4}) we obtain
\begin{eqnarray}
M^{(1)}_{\nu}(0,h)&=&\frac{c^{\nu}_0}{c^{\nu}_4}
\bigg[\frac{c^{\nu}_0}{c^{\nu}_0}
J_{-2}J_{\nu+2}-\frac{c^{\nu}_2}{c^{\nu}_0}J_{-1}J_{\nu+3}\nonumber\\
&-&\frac{c^{\nu}_{-2}}{c^{\nu}_0}J_{-3}J_{\nu+1}
+\frac{c^{\nu}_{4}}{c^{\nu}_0}J_0J_{\nu+4}+
\frac{c^{\nu}_{-4}}{c^{\nu}_0}J_{-4}J_{\nu}+\cdot\cdot\cdot\bigg]
\label{B.11}
\end{eqnarray}
In lowest orders of $h^2$ this is
$$
M^{(1)}_{\nu}(0,h)=\frac{c^{\nu}_0}{c^{\nu}_4}J_{-2}J_{\nu+2}
+\frac{c^{\nu}_{-4}}{c^{\nu}_4}J_{-4}J_{\nu}+\cdot\cdot\cdot
$$
which when evaluated in lowest orders gives
$$
\frac{3.2^7}{h^4}\frac{(h/2)^{\nu+4}}{\nu!}\bigg[
\frac{1}{2(\nu+1)(\nu+2)}-\frac{3}{(2+i\sqrt{5})4.3.2}\bigg]
$$
which reduces to
$$
\frac{1}{2}(h/2)^2
\bigg[\frac{-1+i\sqrt{5}}{2+i\sqrt{5}}\bigg]
$$
in agreement with the previous result, i.e. eq.(\ref{B.9}).

Next we come to  $M^{(1)}_{-\nu}(0,h)$. Again we use first the method
with $r=0$ in eq. (\ref{B.4}).We have
\begin{eqnarray}
M^{(1)}_{-\nu}(0,h)&=& J_0(h)J_{-\nu}(h)-\frac{c^{-\nu}_2(h^2)}{c^{-\nu}_0(h^2)}
J_1(h)J_{-\nu+1}(h)
-\frac{c^{-\nu}_{-2}(h^2)}{c^{-\nu}_0(h^2)}J_{-1}(h)J_{-\nu-1}(h)\nonumber\\
&+&\frac{c^{-\nu}_4(h^2)}{c^{-\nu}_0(h^2)}J_2(h)J_{-\nu+2}(h)
+\frac{c^{-\nu}_{-4}(h^2)}{c^{-\nu}_0(h^2)}J_{-2}(h)J_{-\nu-2}(h)+\cdot\cdot\cdot
\label{B.12}
\end{eqnarray}
In lowest orders of $h^2$ this is
$$
M^{(1)}_{-\nu}(0,h)= J_0(h)J_{-\nu}(h)
+\frac{c^{-\nu}_{4}(h^2)}{c^{-\nu}_0(h^2)}J_{2}(h)J_{-\nu+2}(h)
$$
which when evaluated in lowest orders of $h^2$ implies (with the
help of the power expansion of the Bessel function $J_{\mu}(2h)$)
$$
1.\bigg(\frac{(h/2)^{-\nu}}{(-\nu)!}
+O(h^4)+\frac{(h/2)^{-\nu+4}}{2!(-\nu+2)!}\bigg)
+\frac{1}{2}\frac{c^{-\nu}_4}{c^{-\nu}_0}(h/2)^{4-\nu}
$$
and so
$$
(h/2)^2\bigg(\frac{1}{2}+\frac{(1-4i\sqrt{5})}{6(2+i\sqrt{5})}
\bigg)=+\frac{1}{2}(h/2)^2
-\frac{1}{6}(h/2)^2(2+i\sqrt{5})
$$
It follows that
\begin{equation}
M^{(1)}_{-\nu}(0,h)=\frac{1}{6}(h/2)^2
(1-i\sqrt{5}) 
\label{B.13}
\end{equation}

If we set $r=2$ in eq. (\ref{B.4}) and evaluate $M^{(1)}_{-\nu}(0,h)$,
we obtain
\begin{eqnarray}
M^{(1)}_{-\nu}(0,h)&=&\frac{c^{-\nu}_0}{c^{-\nu}_4}
\bigg[
J_{-2}J_{-\nu+2}-\frac{c^{-\nu}_2}{c^{-\nu}_0}J_{-1}J_{-\nu+3}\nonumber\\
&-&\frac{c^{-\nu}_{-2}}{c^{-\nu}_0}J_{-3}J_{-\nu+1}
+\frac{c^{-\nu}_{4}}{c^{-\nu}_0}J_0J_{-\nu+4}+
\frac{c^{-\nu}_{-4}}{c^{-\nu}_0}J_{-4}J_{-\nu}+\cdot\cdot\cdot\bigg]
\label{B.14}
\end{eqnarray}
In lowest orders of $h^2$ this is
$$\frac{c^{-\nu}_0}{c^{-\nu}_4}
\bigg[
J_{-2}J_{-\nu+2}
+\frac{c^{-\nu}_{4}}{c^{-\nu}_0}J_0J_{-\nu+4}
+\cdot\cdot\cdot\bigg]
$$
Evaluating this as before we obtain in leading orders
\begin{equation}
M^{(1)}_{-\nu}(0,h)
=-\bigg(\frac{2+i\sqrt{5}}{3}\bigg)\bigg[\frac{1}{2}(h/2)^2.1
-\bigg(\frac{3}{2+i\sqrt{5}}\bigg).1.\frac{(h/2)^{-\nu+4}}{(-\nu+4)!}\bigg]
\label{B.15}
\end{equation}
Hence in leading order
\begin{equation}
M^{(1)}_{-\nu}(0,h)
=(h/2)^2\bigg[\frac{1-i\sqrt{5}}{6}\bigg]
\label{B.16}
\end{equation}
which is seen to be in agreement with eq. (\ref{B.13}).
With Mathematica we obtain higher order terms, i.e.
\begin{eqnarray}
M_\nu ^{(1)}(0,h)
&=&\frac{1+i\sqrt{5}}{6}\bigg(\frac{h}{2}\bigg)^2
+\frac{1+i\sqrt{5}}{9}\bigg(\frac{h}{2}\bigg)^4\nonumber\\
&+&
\frac 1{2160}\left( -290+151i\sqrt{5}+120\left(
5-i\sqrt{5}\right) \left(
C+\ln \frac h2\right) \right)\bigg(\frac{h}{2}\bigg)^6 \nonumber\\
&+&\frac 1{3240}\left( -514-73i\sqrt{5}+120\left(
5-i\sqrt{5}\right)
\left( C+\ln \frac h2\right) \right)\bigg(\frac{h}{2}\bigg)^8,
\label{B.17}
\end{eqnarray}

\begin{eqnarray}
M_{-\nu }^{(1)}(0,h)
&=&\frac{1-i\sqrt{5}}{6}\bigg(\frac{h}{2}\bigg)^2
+\frac{1-i\sqrt{5}}{9}\bigg(\frac{h}{2}\bigg)^4\nonumber\\
&+&
\frac 1{2160}\left( -290-151i\sqrt{5}+120\left(
5+i\sqrt{5}\right) \left(
C+\ln \frac h2\right) \right)\bigg(\frac{h}{2}\bigg)^6\nonumber\\
&+&\frac 1{3240}\left( -514+73i\sqrt{5}+120\left(
5+i\sqrt{5}\right)
\left( C+\ln \frac h2\right) \right)\bigg(\frac{h}{2}\bigg)^8
\label{B.18}
\end{eqnarray}
With these results we can evaluate $\alpha_{\nu}$ and $ \alpha_{-\nu}$.
Thus again in the dominant approximation
\begin{equation} 
\alpha_{\nu}=Me_{\nu}(0,h)/M^{(1)}_{\nu}(0,h)=c^{\nu}_0/\bigg[\frac{1}{2}
(h/2)^2\bigg]
\label{B.19}
\end{equation}
and 
\begin{equation}
\alpha_{-\nu}=Me_{-\nu}(0,h)/M^{(1)}_{-\nu}(0,h)=-c^{-\nu}_0
.\bigg(\frac{3}{2+i\sqrt{5}}\bigg)/\bigg[\frac{1}{2}
(h/2)^2\bigg]
\label{B.20}
\end{equation}
It follows that
\begin{eqnarray}
R &=&\frac{M_{-\nu }^{(1)}(0,h)}{M_\nu
^{(1)}(0,h)}\nonumber\\
&=&-\frac{i+\sqrt{5}}{-i+%
\sqrt{5}}+\frac{-49+80\left( C+\ln \frac h2\right)
}{128\sqrt{5}(2i+\sqrt{5})%
}h^4 \nonumber\\
&+&\frac{\left( -49+80\left( C+\ln \frac h2\right)
\right) \left( 151i-58%
\sqrt{5}+120\left( -i+\sqrt{5}\right) \left( C+\ln
\frac h2\right) \right) }{%
737280(-7i+\sqrt{5})}h^8\nonumber\\
&=&-\frac{2+i\sqrt{5}}3+\frac{\left(
5-2i\sqrt{5}\right) \left(-49+80\left(
C+\ln \frac h2\right) \right) }{360}\left( \frac
h2\right) ^4 \nonumber\\
&+&\frac{1}{51840}\bigg\{49\left(449+85i\sqrt{5}\right) -80\left(
743+232i\sqrt{5}\right)
\left(C+\ln \frac h2\right)\nonumber\\
&+&19200\left(2+i\sqrt{5}\right)\left(C
+\ln{\frac h2}\right)^2\bigg\}\left(\frac h2\right)^8
\label{B.21}
\end{eqnarray}
with $c^{\nu}_0=c^{-\nu}_0=1$ (cf.MS, p. 122, eq. (39)).  

We can now identify our quantities with those of ref.\cite{1}. Comparison
with the Dougall coefficients evaluated previously implies
\begin{equation}
\alpha_{\nu}=1/{\phi(\nu/2)}, \;\;\;\alpha_{-\nu}=1/{\phi(-\nu/2)} 
\label{B.22}
\end{equation}
and so in the notation of ref.\cite{1} 
\begin{equation}
R = {\phi(-\nu/2)}/{\phi(\nu/2)}
\label{B.23}
\end{equation}
A remarkable feature of the expression (\ref{B.21}) is its unit modulus,
as was also observed in ref.\cite{1}. It means that $R$ is a pure phase
factor 
\begin{equation}
R = e^{i\pi\gamma}, RR^*=1
\label{B.24}
\end{equation}
When and why this behaviour occurs is discussed
at the end of Appendix A.

\section{The quantity $R$ in the general case}
\label{sec:VIII}

In the general case, or for $s=l+2$ sufficiently large so that no
problems with singularities arise, we can evaluate $R$ and so
$M^{(1)}_{\pm \nu}(0,h)$ and $Me_{\pm\nu}(0,h)$ by
simply using the power series expansions of $\nu$ and 
the standard Mathieu coefficients and, of course, the power series
expansion of Bessel functions $J_{\mu}(2h)$. One then
obtains
\begin{eqnarray}
M^{(1)}_{\pm \nu}(0,h)&=&\frac{1}{(\pm\nu) !}\bigg(\frac{h}{2}\bigg)^{\pm\nu}
\bigg[1+\frac{2}{\nu^2 -1}\bigg(\frac{h}{2}\bigg)^2\mp\frac{2(\nu^2\mp 3\nu -7)}
{(\nu\pm 1)^2(\nu\mp 1)(\nu^2 -4)}\bigg(\frac{h}{2}\bigg)^4\nonumber\\
&\mp& \frac{4(\nu^4\mp 11\nu^3-2\nu^2\pm 59\nu -23)}
{(\nu \pm 1)^2(\nu\mp 1)^3(\nu^2 -4)(\nu^2 -9)}\bigg(\frac{h}{2}\bigg)^6
+\cdot\cdot\cdot\bigg]
\label{D.1}
\end{eqnarray}
This implies for $R$
\begin{eqnarray}
R&=&\frac{\nu !}{(-\nu)!\bigg(\frac {h}{2}\bigg)^{2\nu}}\cdot
\frac{\bigg[1+\frac{2}{\nu^2 -1}\bigg(\frac{h}{2}\bigg)^2 +
\frac{2(\nu^2+ 3\nu -7)}
{(\nu- 1)^2(\nu+ 1)(\nu^2 -4)}\bigg(\frac{h}{2}\bigg)^4
+\cdot\cdot\cdot\bigg]}
{\bigg[1+\frac{2}{\nu^2 -1}\bigg(\frac{h}{2}\bigg)^2-\frac{2(\nu^2- 3\nu -7)}
{(\nu+ 1)^2(\nu -1)(\nu^2 -4)}\bigg(\frac{h}{2}\bigg)^4+\cdot\cdot\cdot \bigg]}
\nonumber\\
&=&\frac{\nu !(\nu-1)!\sin\pi\nu}{\pi\bigg(\frac{h}{2}\bigg)^{2\nu}}
\cdot\bigg[1+\frac{4\nu}{(\nu^2-1)^2}\bigg(\frac{h}{2}\bigg)^4\nonumber\\
&+&\frac{2\nu(4\nu^5+15\nu^4-32\nu^3-12\nu^2+64\nu-111)}
{(\nu^2-1)^4(\nu^2-4)^2}\bigg(\frac{h}{2}\bigg)^8+\cdot\cdot\cdot \bigg]
\label{D.2}
\end{eqnarray}
The first few terms of the expansion of the function $Me_{\nu}(0,h)$
which is needed for comparison with
the results of ref. \cite{1} have been obtained in the previous section.

From eq.(\ref{D.2}) we extract for later reference
\begin{equation}
\bigg(\frac{\sin\pi\nu}{R}\bigg)^2=\frac{\pi^2\bigg(\frac{h}{2}\bigg)^{4\nu}}
{\bigg\{\nu!(\nu-1)!\bigg\}^2\bigg[1+
\frac{4\nu}{(\nu^2-1)^2}\bigg(\frac{h}{2}\bigg)^4 + \cdot\cdot\cdot \bigg]^2}
\label{D.3}
\end{equation}
This expansion will be used below in the low order
approximation of the absorptivity for higher partial waves.

\section{Calculation of the absorptivity}
\label{sec:IX}

We consider the absorptivity in a general case, and hence allow
for complex Floquet exponents $\nu$, which we set
\begin{equation}
\nu=n+i(\alpha+i\beta)=(n-\beta)+i\alpha
\label{E.1}
\end{equation}
where $n=2,3,4,...$ and $\alpha$ and $\beta$ are real and of $O(h^4)$. 
In evaluating the $S$--matrix for small $h^4$ one has to be
careful to make the expansions in the appropriate way. Thus we
write $SS^{\star}$
\begin{equation}
SS^{\star}=\frac{(1-\frac{1}{R^2})(1-\frac{1}{{R^{\star}}^2}) }
{(e^{i\pi\nu}-\frac{e^{-i\pi\nu}}{R^2})(e^{-i\pi\nu^{\star}}
-\frac{e^{i\pi\nu^{\star}}}{{R^{\star}}^2})}
\label{E.2}
\end{equation}
which can be rewritten
\begin{equation}
SS^{\star}=
\frac
{e^{2\pi\alpha}(1-\frac{1}{R^2})(1-\frac{1}{{R^{\star}}^2})}
{\bigg[1-\bigg\{\cos2\pi\beta.(\frac{1}{R^2}+\frac{1}{{R^{\star}}^2})
.e^{2\pi\alpha}
+i\sin2\pi\beta.(\frac{1}{R^2}-\frac{1}{{R^{\star}}^2}).e^{2\pi\alpha}
-\frac{e^{4\pi\alpha}}{R^2{R^{\star}}^2}\bigg\}\bigg]}
\label{E.3}
\end{equation}
Here we set
\begin{equation}
e^{2i\pi\beta}\equiv 1+if, \;\;\; e^{2\pi\alpha}\equiv 1+g
\label{E.4}
\end{equation}
where $f$ is complex and $g$ is real (in the $S$--wave case
$g=-\frac{2\sqrt{5}\pi}{3}(h/2)^4+O(h^8)$).  Then
\begin{equation}
\cos 2\pi\beta=1-\Im f, \Im f\approx \frac{1}{2}(2\pi\beta)^2\approx 
\frac{1}{2}(\sin2\pi\beta)^2 
\label{E.5}
\end{equation}
and
\begin{equation}
\sin 2\pi\beta=\Re f
\label{E.6}
\end{equation}
Then
\begin{eqnarray}
SS^{\star}&=&(1+g)
\bigg[1\nonumber\\
&-&\frac{\bigg\{g(\frac{1}{R^2}
+\frac{1}{{R^{\star}}^2}-\frac{2+g}{R^2{R^{\star}}^2})
-\Im f.(1+g)(\frac{1}{R^2}+\frac{1}{{R^{\star}}^2})
+(1+g).\Re f.i(\frac{1}{R^2}-\frac{1}{{R^{\star}}^2})\bigg\}}
{(1-\frac{1}{R^2}) (1-\frac{1}{{R^{\star}}^2})}\bigg]^{-1}
\label{E.7}
\end{eqnarray}

We now consider two limiting cases.

(i) $\alpha\rightarrow 0$ implying $g\rightarrow 0.$

\noindent
In this case $R=R^{\star}$ and so $1/R^2\simeq O(h^4)$.
This is the case of real Floquet exponents and so
excludes the case of $S$--waves. Here
\begin{equation}
SS^{\star}=\frac{1}{1+\frac{\Im f.{\frac{2}{R^2}}}{(1-\frac{1}{R^2})^2}}
\simeq 1-\frac{2\Im f}{\frac{R^2}{(1-\frac{1}{R^2})^2}}
\simeq 1-\frac{4\bigg(\frac{\sin\pi\beta}{R}\bigg)^2}{(1-\frac{1}{R^2})^2}
\label{E.8}
\end{equation}
since
$$
\Im f\simeq \frac{1}{2}(\sin2\pi\beta)^2\simeq 2\sin^2\pi\beta.
$$
The absorptivity $A$ is therefore given by
\begin{equation}
A=1-SS^{\star}\approx \frac{4\bigg(\frac{\sin\pi\beta}{R}\bigg)^2}
{(1-\frac{1}{R^2})^2}\approx 4\bigg(\frac{\sin\pi\beta}{R}\bigg)^2
\label{E.9}
\end{equation}
With the help of eq.(\ref{D.3}) this can be written
\begin{eqnarray}
A&\approx &\frac{4\pi^2\bigg(\frac{h}{2}\bigg)^{4\nu}}
{\bigg\{\nu!(\nu-1)!\bigg\}^2}\cdot
\bigg[1+\frac{4\nu}{(\nu^2-1)^2}
\bigg(\frac{h}{2}\bigg)^4+O(h^8)\bigg]^{-2}\nonumber\\
&=&\frac{4\pi^2\bigg(\frac{h}{2}\bigg)^{4(l+2)}}{\bigg\{(l+1)!(l+2)!\bigg\}^2}
\cdot\bigg[1+\frac{4\nu}{(\nu^2-1)^2}
\bigg(\frac{h}{2}\bigg)^4+ O(h^8)\bigg]^{-2}
\label{E.10}
\end{eqnarray}
in agreement with ref.\cite{1}. This
can be easily evaluated (e.g. with Maple), e.g.
already in the case of $P$--waves (i.e. in spite of singularities
in higher order terms here omitted) and yields in this case
\begin{equation}
A=\frac{\pi^2 h^{12}}{3^2 2^{14}}\bigg[1+\bigg\{\frac{53}{1152}
-\frac{1}{24}\log\bigg(\frac{he^{\gamma}}{2}\bigg)\bigg\}h^4\bigg]
\label{E.11}
\end{equation}
where $\gamma$ is the Euler constant (also written $C$).
 This result agrees with
the result in ref.\cite{1}. We observe, in particular, that logarithmic
energy contributions arise in the expansion.
The formula (\ref{E.10}), of course, does
not apply in the case of $S$--waves.

(ii) $\beta\rightarrow 0$ implying $f\rightarrow 0.$

\noindent
In this case $RR^{\star}=1$, so that $|R|\sim O(h^0)$, and we
cannot expand as in the previous case. However, $g\approx O(h^4)$,
so that we can expand in powers of $g$.  Thus
\begin{eqnarray}
SS^{\star}&=&\frac{1+g}{1+g\bigg\{1+\frac{g}{2-R^2-{R^{\star}}^2}\bigg\}}
\nonumber\\
&= & 1-\frac{g^2}{2-R^2-{R^{\star}}^2}+O(g^3)\nonumber\\
&=& 1+O(h^8)
\label{E.11.1}
\end{eqnarray}
This is the case of complex Floquet exponents as in the $S$--wave case.
In this case
\begin{equation}
SS^{\star}=1-\frac{9g^2}{20}+O(g^3)
\label{E.12}
\end{equation}
and so
\begin{equation}
A=\frac{9g^2}{20}+O(g^3)=\pi^2(h/2)^8+\frac{2\pi ^2}9 \left(7-12(\gamma+\ln
\frac h2) \right) (h/2) ^{12}+O(h^{16})
\label{E.13}
\end{equation}
in agreement with ref.\cite{1} and the rough lowest order
calculation of ref.\cite{11.2}.
We can also compute for this particularly interesting $S$--wave case
the amplitudes of the reflected, transmitted and incident waves
$A_r, A_t, A_i$ defined in Appendix A. One finds
\begin{eqnarray}
A_r&=& R-\frac{1}{R}= -\frac{2i\sqrt{5}}{3}+O(h^4),\nonumber\\
A_t&=& 2i\sin\nu\pi=\frac{2\sqrt{5}}{3}\pi(h/2)^4+O(h^8)\nonumber\\
A_i&=&Re^{i\nu\pi}-\frac{1}{Re^{i\nu\pi}}
=-\frac{2i\sqrt{5}}{3}-\frac{4\sqrt{5}\pi}{9}(h/2)^4+O(h^8)
\label{E.14}
\end{eqnarray}
We observe that in the limit $h^4\rightarrow 0$
 we have $A_t=0$ and $A_r=A_i$, i.e.
there is only reflection of the fluctuation or disturbance around the
$D$--brane like reflection from a wall and no transmission, which can be
interpreted as a vanishing of the disturbance
on the brane (implying a Dirichlet boundary
condition).  On the basis of the analogy with the
case of the open fundamental string between brane and antibrane in
Born--Infeld theory we can expect that as the energy increases,
transmission (i.e. absorption) sets in and becomes the dominant
effect at high energies.  This is similar to what one
finds in quantum mechanics of a potential well of depth $-V_o$ \cite{14}.
There the properly normalised transmission and reflection
coefficients $T(E), R(E)$, where $E$ is the energy, have
the behaviour $T(E)\rightarrow 0, R(E)\rightarrow 1$
 as $E\rightarrow 0$, but $T(E)\rightarrow 1$ and
$R(E) \rightarrow 0$ as $E\rightarrow\infty$.
The high energy behaviour of the effect considered here
can presumably be investigated with the help of large--$h$
asymptotic expansions of modified Mathieu functions
which we expect to be formally (i.e. apart from 
sign and complex $i$ changes) similar to those of periodic
Mathieu functions with a parameter $q$ defined as the
solution of
\begin{equation}
(l+2)^2= -2h^2+2hq+O(h^0).
\label{E.15}
\end{equation}
The Floquet exponent is then given by \cite{10}(p.210),\cite{lang}
\begin{equation}
\cos\pi\nu + 1 =\frac{\pi e^{4h}}{(8h)^{q/2}}\left[\frac
{1+3(q^2+1)/64h}{\Gamma((1-q)/4)\Gamma((3-q)/4)}
+O\left(\frac{\log h}{h^2}\right)\right].
\label{E.16}
\end{equation}
One observes that again logarithmic contributions 
in the energy appear.

\section{Conclusions}
\label{sec: X}
In the above we considered the impingement of a massless scalar
field on a $D3$ brane in 10 dimensions and calculated the 
$S$ matrix and partial wave absorption and reflection
amplitudes and rates for this process. Instead of coefficients
introduced by Dougall for the expansion of the modified Mathieu
functions involved, we used (in the low energy
domain) rapidly convergent series in terms of products of
Bessel functions. We demonstrated that the Mathieu
function coefficients are such that many different expansions
in terms of products of Bessel functions all yield the
same low energy power series for the modified Mathieu functions
of the first kind. We think, this is the best
way to evaluate the absorption rates of the
problem in the low energy domain. The leading term matching
procedures of refs. \cite{11.1,11.2} maybe select
dominant terms of the expansions considered here.
Since the metric considered is extremal,
one can visualise the absorption of the partial waves of the
scalar field as absorption into the brane or black hole
with vanishing event horizon (examples with
nonvanishing horizon have for instance been treated
in \cite{korea}). Since several other string models lead
also to the modified Mathieu equation in analogous contexts,
the above considerations, which have definite
advantages over those involving the
coefficients of Dougall, may be of wider interest.

\vskip 1cm
\noindent
{\bf Acknowledgement}:R.M. and Y.Z. acknowledge support as
Fellows of the A. von Humboldt Foundation. J.--Q. L. acknowledges support
by the Deutsche Forschungsgemeinschaft in the framework of the
agreement with the Ministry of Education of China.
H.J.W. M.--K.
is indebted to A. Hashimoto for correspondence.

\newpage

\begin{appendix}{\centerline{\bf Appendix A}}

\setcounter{equation}{0}
\renewcommand{\theequation}{A.\arabic{equation}}
Here we recapitulate the main steps of the derivation of the
$S$--matrix. We follow ref.\cite{6}, but instead of repeating
the steps there, we emphasize those which have not been
written out explicitly there. For ease of
comparison we consider the repulsive potential
which means simply that the
considerations below employ
coupling $g$ as in ref.\cite{6}
instead of $g_0$ used above.  The two cases
are trivially related through
\begin{equation}
g_0=ig
\end{equation}
\label{A.1}
In the repulsive case we have a regular solution
$y_{reg}$ of eq.(\ref{5})at $r=0$, i.e. one proportional to
$exp(-g/r)$.  The variable of the cylindrical
functions involved in $M^{(j)}(z,h)$ is $\omega=2h\cosh z =
(ig/r+kr)$. Thus in leading order for small $h^2$ and
$r\rightarrow 0  (z\rightarrow -\infty)$ we can write
\begin{equation}
y_{reg}=r^{1/2}M^{(3)}_{\nu}(z,h)
\stackrel{\Re z\rightarrow -\infty}{\simeq}
r^{1/2}\bigg[H^{(1)}_{\nu}(\omega)+O(h^2)\bigg]
\stackrel{r\rightarrow 0}{\simeq}
\bigg(\frac{2}{\pi g}\bigg)^{\frac{1}{2}}e^{-\frac{g}{r}}
e^{-i(\nu+1)\frac{\pi}{2}}\bigg[1+O(h^2)\bigg]
\end{equation}
\label{A.2}
If we let $Re z\rightarrow -\infty$ here and then replace 
$z$ by$-z$, the solution has the asymptotic
behaviour
$e^{ikr}$.
The series expansion defining the Bessel function $J_{\nu}$
has the following important property for integers $n$
\begin{equation}
J_{\nu}(2h\cosh(z+in\pi))=exp(in\nu\pi)J_{\nu}(2h\cosh z)
\end{equation}
\label{A.3}
so that
\begin{equation}
M^{(1)}_{\nu}(z+in\pi,h) = exp(in\nu\pi)M^{(1)}_{\nu}(z,h).
\end{equation}
\label{A.4}
Since for $ Me_{\nu}(z,h)$ also
\begin{equation}
 Me_{\nu}(z+in\pi,h)=exp(in\nu\pi) Me_{\nu}(z,h)
\end{equation}
\label{A.5}
we have the proportionality
\begin{equation}
 Me_{\nu}(z,h)=\alpha_{\nu}(h)M^{(1)}_{\nu}(z,h)
\label{A.6}
\end{equation}
with (e.g.)
\begin{equation}
\alpha_{\nu}(h)= \frac{Me_{\nu}(0,h)}{M^{(1)}_{\nu}(0,h)}
\end{equation}
\label{A.7}
As mentioned earlier, expansions in terms of cylindrical functions like
(\ref{B.2}) converge uniformly only in domains $|\cosh z| >1$, whereas the
expansion (\ref{a.1}) of  $Me_{\nu}(z,h)$ converges for all
finite complex values of $z$. Hence we match $M^{(3)}_{\nu}(z,h)$
in the domain $\Re z<0 $ to a linear combination of
$M^{(3)}_{\nu}(z,h)$ and $M^{(4)}_{\nu}(z,h)$
in the domain $\Re z>0 $ by matching  both to
a combination of  $Me_{\nu}(z,h)$
and $Me_{-\nu}(z,h)$ in the intermediate domain.
We have
\begin{equation}
z=\log\sqrt{k/g}r \stackrel{-}{(+)} i\pi/4.
\label{A.8}
\end{equation}
In the domain of $r$ close to zero we set
\begin{eqnarray}
r^{1/2}M^{(3)}_{\nu}&=&r^{1/2}\bigg(\alpha Me_{\nu}
+\beta Me_{-\nu}\bigg)\nonumber\\
\frac{d}{dr}\bigg(r^{1/2}M^{(3)}_{\nu}\bigg)&=&\alpha\frac{d}{dr}\bigg(r^{1/2}
Me_{\nu}\bigg)
+\beta\frac{d}{dr}\bigg(r^{1/2}Me_{-\nu}\bigg)
\label{A.9}
\end{eqnarray}
where $\alpha$ and $\beta$ have to be determined. In the domain of large
$r$ we set, with constants $\alpha^{\prime}, \beta^{\prime}, A, B$, which
have to be determined
\begin{eqnarray}
r^{1/2}\bigg(\alpha^{\prime} Me_{\nu}
+\beta^{\prime} Me_{-\nu}\bigg)&=&r^{1/2}\bigg(AM^{(3)}_{\nu}
+BM^{(4)}_{\nu}\bigg),\nonumber\\
\alpha^{\prime}\frac{d}{dr}\bigg(r^{1/2}Me_{\nu}\bigg)
+\beta^{\prime}\frac{d}{dr}\bigg(r^{1/2}Me_{-\nu}\bigg)&=&A\frac{d}{dr}
\bigg(r^{1/2}M^{(3)}_{\nu}\bigg)
+B\frac{d}{dr}\bigg(r^{1/2}M^{(4)}_{\nu}\bigg).
\label{A.10}
\end{eqnarray}
We match the $Me_{\nu}, Me_{-\nu}$ combination (variable $z$) on the left
to that on the right (variable $-z$) at $\Re z = 0 (r =\sqrt{g/k})$, so that
\begin{eqnarray}
r^{1/2}\bigg(\alpha Me_{\nu}
+\beta Me_{-\nu}\bigg)_{z=+i\pi/4}&=&r^{1/2}\bigg(\alpha^{\prime} Me_{\nu}
+\beta^{\prime} Me_{-\nu}\bigg)_{z=-i\pi/4},\nonumber\\
\bigg[\alpha\frac{d}{dr}\bigg(r^{1/2}Me_{\nu}\bigg)+\beta\frac{d}{dr}
\bigg(r^{1/2}Me_{-\nu}\bigg)\bigg]_{z=+i\pi/4}
&=&\bigg[\alpha^{\prime}\frac{d}{dr}\bigg(r^{1/2}Me_{\nu}\bigg)
+\beta^{\prime}\frac{d}{dr}\bigg(r^{1/2}Me_{-\nu}\bigg)\bigg]_{z=-i\pi/4}
\label{A.11}
\end{eqnarray}
Since $Me_{\nu}(z)=Me_{-\nu}(-z)$ and at $r=\sqrt{g/k}, z=\pm i\pi/4$,
 also
$$
\frac{d}{dr}=\mp\bigg(\frac{k}{g}\bigg)^{1/2}\frac{d}{dz},
$$
the latter become
\begin{eqnarray}
\alpha Me_{-\nu}+\beta Me_{\nu}&=&\alpha^{\prime}Me_{\nu}
+\beta^{\prime}Me_{-\nu}
\nonumber\\
\alpha\frac{d}{dz} Me_{-\nu}+\beta\frac{d}{dz}Me_{\nu}&=&\alpha^{\prime}
\frac{d}{dz}Me_{\nu}+\beta^{\prime}\frac{d}{dz}Me_{-\nu}
\label{A.12}
\end{eqnarray}
From these equations we obtain immediately
\begin{equation}
\alpha^{\prime}=\beta, \;\;\; \beta^{\prime}=\alpha
\label{A.13}
\end{equation}
From eqs.(\ref{A.9}) we obtain ($W$ meaning Wronskian)
\begin{equation}
\alpha=\frac{W[M^{(3)}_{\nu}, Me_{-\nu}]}{W[Me_{\nu},
Me_{-\nu}]}, \;\;\;
\beta=-\frac{W[M^{(3)}_{\nu}, Me_{\nu}]}{W[Me_{\nu},
Me_{-\nu}]},
\label{A.14}
\end{equation}
From (\ref{A.10}) we obtain similarly
\begin{eqnarray}
A&=&\frac{-W[M^{(3)}_{\nu}, Me_{\nu}]W[Me_{\nu}, M^{(4)}_{\nu}]
+W[M^{(3)}_{\nu}, Me_{-\nu}]W[Me_{-\nu}, M^{(4)}_{\nu}]}
{W[M^{(3)}_{\nu}, M^{(4)}_{\nu}]W[Me_{\nu}, Me_{-\nu}]},\nonumber\\
B&=&\frac{W[M^{(3)}_{\nu}, Me_{\nu}]W[Me_{\nu}, M^{(3)}_{\nu}]
-W[M^{(3)}_{\nu}, Me_{-\nu}]W[Me_{-\nu}, M^{(3)}_{\nu}]}
{W[M^{(3)}_{\nu}, M^{(4)}_{\nu}]W[Me_{\nu}, Me_{-\nu}]}
\label{A.15}
\end{eqnarray}
We now use eq.(\ref{A.6}) and Wronskians $W[M^{(i)}_{\nu},M^{(j)}_{\nu}]
\equiv[i,j]$ given in ref.\cite{10},
 i.e.
\begin{equation}
[3,4]=-\frac{4i}{\pi},\;\;[1,3]=-[1,4]=\frac{2i}{\pi}
\label{A.16}
\end{equation}
and the circuit relation (\cite{10}, p. 169)
\begin{equation}
M^{(1)}_{-\nu}=e^{i\nu\pi}M^{(1)}_{\nu}-i\sin\nu\pi M^{(4)}_{\nu}.
\label{A.17}
\end{equation}
Then
\begin{eqnarray}
W[Me_{\nu}, Me_{-\nu}]&=&-\frac{2\sin\nu\pi}{\pi}\alpha_{\nu}\alpha_{-\nu},
\nonumber\\
W[Me_{-\nu},M^{(3)}_{\nu}]=\frac{2i}{\pi}e^{-i\nu\pi}\alpha_{-\nu}&,&
W[Me_{-\nu},M^{(4)}_{\nu}]=-\frac{2i}{\pi}e^{i\nu\pi}\alpha_{-\nu},
\label{A.18}
\end{eqnarray}
With these expressions $A$ and $B$ are found to be
\begin{equation}
A=\frac{1}{2i\sin\nu\pi}\bigg({\frac{\alpha_{\nu}}{\alpha_{-\nu}}
-\frac{\alpha_{-\nu}}{\alpha_{\nu}}}\bigg)
,\;\;\;
B=\frac{1}{2i\sin\nu\pi}\bigg({\frac{\alpha_{\nu}}{\alpha_{-\nu}}-
e^{-2i\nu\pi}\frac{\alpha_{-\nu}}{\alpha_{\nu}}}\bigg)
\label{A.19}
\end{equation}
The regular solution thus continued to $r=\infty$ is then
\begin{eqnarray}
y_{reg}&\simeq& r^{1/2}\bigg[AM^{(3)}(z,h)+BM^{(4)}(z,h)\bigg]\nonumber\\
&\simeq&\bigg(\frac{2}{k\pi}\bigg)^{1/2}
\bigg\{A e^{ikr}e^{-i(\nu+\frac{1}{2})\frac{\pi}{2}}
+e^{-i\frac{\pi}{2}}Be^{-ikr}e^{i(\nu+\frac{1}{2})\frac{\pi}{2}}\bigg\}
\label{A.20}
\end{eqnarray}
In terms of the variable $z$ and with $R\equiv \alpha_{\nu}/\alpha_{-\nu}$
this can be written
\begin{eqnarray}
&\;&\bigg(\frac{2r}{2h\pi\cosh z}\bigg)^{\frac{1}{2}} e^{-i(\nu+\frac{1}{2})
\frac{\pi}{2}} \bigg\{2i\sin\nu\pi\bigg\} e^{2ih\cosh z}\nonumber\\
&\;&\stackrel{\Re z\rightarrow \infty}{\simeq}
\bigg(\frac{2r}{2h\pi\cosh z}\bigg)^{\frac{1}{2}} e^{-i(\nu+\frac{1}{2})
\frac{\pi}{2}}\bigg\{(R-\frac{1}{R})e^{2ih\cosh z}\nonumber\\
&+&i(Re^{i\nu\pi}-
\frac{e^{-i\nu\pi}}{R})e^{-2ih\cosh z}\bigg\}
\label{A.21}
\end{eqnarray}
If we take $A_i=(Re^{i\nu\pi}-\frac{e^{-i\nu\pi}}{R})$ as the
amplitude of the incident wave, the amplitudes $A_r$ and $A_t$
of the reflected and transmitted waves
are $A_r=R-\frac{1}{R}$ and $A_t= 2i\sin\nu\pi$
respectively in agreement with ref.\cite{1}. With
the definition of the $S$--matrix in the partial wave
expansion of the
scattering amplitude $f(\theta)$, with $x=\cos\theta$,
 for $n$ space dimensions (here we have $n=6$)
given by\cite{12}
\begin{equation}
e^{ikx}+f(\theta)\frac{e^{ikr}}{r^{(n-1)/2}}\simeq 
\frac{1}{2(ikr)^{(n-1)/2}}\sum^{\infty}_{l=0}\bigg\{Se^{ikr}
+(-1)^li^{n-1}e^{-ikr}\bigg\}{\tilde {\cal P}}_l(\cos\theta)
\label{A.22}
\end{equation}
where
$$
{\tilde {\cal P}}_l(\cos\theta)
=\sqrt{\frac{2}{\pi}}2^{n/2-1}\Gamma(n/2-1)(l+\frac{n}{2}-1)C_l(\cos\theta)
$$
and $C_l(\cos\theta)$ is a Gegenbauer polynomial,
we obtain for this
\begin{equation}
S=\frac{R-\frac{1}{R}}{(Re^{i\nu\pi}-\frac{e^{-i\nu\pi}}{R})}.e^{-i\pi l}
\label{A.23}
\end{equation}
It is easy to verify that for $\nu$ real and $R\equiv e^y$ real, 
unitarity is preserved, i.e.
unity minus reflection probability = transmission probability, i.e.
\begin{equation}
1-\frac{|R-\frac{1}{R}|^2}{|Re^{i\nu\pi}-\frac{e^{-i\nu\pi}}{R}|^2}
=\frac{|2\sin\nu\pi|^2}{|Re^{i\nu\pi}-\frac{e^{-i\nu\pi}}{R}|^2}
\label{A.24}
\end{equation}
We observe that this relation remains valid if the real quantity
$R\equiv e^y$ and the pure phase factor $e^{i\pi\nu}$ exchange
their roles, i.e. if $R$ becomes a pure phase factor and
$e^{i\pi\nu}$ a real exponential.  The latter is precisely what happens in
the $S$--wave case of the attractive potential discussed above.

\end{appendix}

\newpage

\begin{appendix}{\centerline {\bf Appendix B}}
\setcounter{equation}{0}
\renewcommand{\theequation}{B.\arabic{equation}}

Below we give the explicit form of the first three terms
of the small--$h^2$ perturbation expansions of
the leading coefficients $c^{-\nu}_{2r}$ of expansions
of modified Mathieu functions in the $S$--wave
case ($l=0$).  The nonleading terms
have been obtained with Mathematica.

\[
\frac{c_2^{-\nu }}{c_0^{-\nu }}=\frac{i+\sqrt{5}}{-2i+\sqrt{5}}\left( \frac
h2\right) ^2+\frac{11(5-2i\sqrt{5})}{2^35(-2i+\sqrt{5})^2}\left( \frac
h2\right) ^6+\frac{(72311i-17746\sqrt{5})}{2^73^35(-2i+\sqrt{5})^3}\left(
\frac h2\right) ^{10}+O(h^{11}),
\]
\[
\frac{c_{-2}^{-\nu }}{c_0^{-\nu }}=-\frac 13\left( \frac h2\right) ^2-\frac{%
3+16i\sqrt{5}}{2^33^3}\left( \frac h2\right) ^6+\frac{5146-391i\sqrt{5}}{%
2^63^55}\left( \frac h2\right) ^{10}+O(h^{11}),
\]
\[
\frac{c_4^{-\nu }}{c_0^{-\nu }}=\frac{i-\sqrt{5}}{i+\sqrt{5}}+\frac{11(5i+%
\sqrt{5})}{2^33^15(i+\sqrt{5})}\left( \frac h2\right) ^4+\frac{2966\sqrt{5}%
-28889i}{2^73^55(2i+\sqrt{5})}\left( \frac h2\right) ^8+O(h^9),
\]
\[
\frac{c_{-4}^{-\nu }}{c_0^{-\nu }}=\frac 1{2^33}\left( \frac h2\right) ^4+%
\frac{18+125i\sqrt{5}}{2^63^35}\left( \frac h2\right) ^8+\frac{1303i\sqrt{5}%
-25774}{2^83^55^2}\left( \frac h2\right) ^{12}+O(h^{13}),
\]

\[
\frac{c_6^{-\nu }}{c_0^{-\nu }}=\frac 1{2+i\sqrt{5}}\left( \frac h2\right)
^2-\frac{290i+49\sqrt{5}}{2^23^35(i+\sqrt{5})}\left( \frac h2\right) ^6+%
\frac{39559i-4597\sqrt{5}}{2^73^55(-i+\sqrt{5})}\left( \frac h2\right)
^{10}+O(h^{11}), 
\]

\[
\frac{c_{-6}^{-\nu }}{c_0^{-\nu }}=-\frac 1{2^33^25}\left( \frac h2\right)
^6-\frac{19+157i\sqrt{5}}{2^63^45^2}\left( \frac h2\right) ^{10}+\frac{%
132035-5159i\sqrt{5}}{2^73^65^37}\left( \frac h2\right) ^{14}+O(h^{15}). 
\]
\end{appendix}

\end{document}